\documentclass[12pt,journal,final,onecolumn]{IEEEtran}

\usepackage[pdftex]{graphicx}
\usepackage[cmex10]{amsmath}
\usepackage{amssymb}
\usepackage{amsthm}
\usepackage{amsfonts}
\usepackage{bm}
\usepackage{cite}
\usepackage[svgnames]{xcolor} % load before pstricks

\graphicspath{{figs/}}

\newcommand{\mc}[1]{\mathcal{#1}}

\newcommand{\mbb}[1]{\mathbb{#1}}

\newcommand{\msf}[1]{\mathsf{#1}}

\newcommand{\DDC}{\text{DDC}}
\newcommand{\DMC}{\text{DMC}}

\newcommand{\defeq}{\mathrel{\triangleq}}
\newcommand{\Pp}{\mathbb{P}}
\newcommand{\E}{\mathbb{E}}
\newcommand{\N}{\mathbb{N}}

\newcommand{\R}{\mathbb{R}}
\newcommand{\Rp}{\R_{+}}

\newcommand{\floor}[1]{\lfloor{#1}\rfloor}
\newcommand{\abs}[1]{\lvert{#1}\rvert}
\newcommand{\card}[1]{\abs{#1}}

\newcommand{\iid}{i.\@i.\@d.\ }

\newcommand{\from}{\colon}

\DeclareMathOperator{\var}{var}

\newtheorem{lemma}{Lemma}

\newtheorem{theorem}[lemma]{Theorem}
\newtheorem{corollary}[lemma]{Corollary}

\theoremstyle{definition}
\newtheorem*{definition}{Definition}

\newtheorem{egdummy}{Example}
\newenvironment{example}[1][]%
{%
    \begin{egdummy}[#1]%
    \upshape%
}%
{%
    \qed%
    \end{egdummy}%
}

\newtheoremstyle{myremark}%
{\topsep}{\topsep}{\normalfont}{\parindent}{\itshape}{:}{ }{}

\theoremstyle{myremark}
\newtheorem*{remark}{Remark}

\newcommand\shortintertext[1]{%
\ifvmode\else\\\@empty\fi
\noalign{%
\penalty0%
\vbox{\mathstrut}%
\penalty10000%
\vskip-\baselineskip
\penalty10000%
\vbox to 0pt{%
\normalbaselines
\ifdim\linewidth=\columnwidth
\else
\parshape\@ne
\@totalleftmargin\linewidth
\fi
\vss
\noindent#1\par}%
\penalty10000%
\vskip-\baselineskip}%
\penalty10000}

\begin{document}

\title{Energy-Efficient Communication in \\ the Presence of Synchronization Errors}

\author{Yu-Chih Huang, Urs Niesen, and Piyush Gupta%
\thanks{This work was supported in part by the Air Force Office of
Scientific Research, Arlington, VA, USA, under grant FA9550-09-1-0317.}%
\thanks{This paper was presented in part at the 2013 IEEE International
Symposium on Information Theory.}%
\thanks{Y.-C.~Huang is with the Department of Communication Engineering,
National Taipei University, New Taipei City, 23741, Taiwan (e-mail:
ychuang@mail.ntpu.edu.tw). U.~Niesen and P.~Gupta were with Bell Labs,
Alcatel-Lucent, Murray Hill, NJ 07974, USA. They are now with the
Qualcomm New Jersey Research Center, Bridgewater, NJ 08807, USA (email:
urs.niesen@ieee.org, p.gupta@ieee.org).}%
}

\maketitle

\begin{abstract}
    Communication systems are traditionally designed to have tight
    transmitter-receiver synchronization. This requirement has
    negligible overhead in the high-SNR regime. However, in many
    applications, such as wireless sensor networks, communication needs
    to happen primarily in the energy-efficient regime of low SNR, where
    requiring tight synchronization can be highly suboptimal.

    In this paper, we model the noisy channel with synchronization
    errors as a
    duplication/{\allowbreak}deletion/{\allowbreak}substitution channel.
    For this channel, we propose a new communication scheme that
    requires only loose transmitter-receiver synchronization. We show
    that the proposed scheme is asymptotically optimal for the Gaussian
    channel with synchronization errors in terms of energy efficiency as
    measured by the rate per unit energy. In the process, we also
    establish that the lack of synchronization causes negligible loss in
    energy efficiency. We further show that, for a general discrete
    memoryless channel with synchronization errors and a general input
    cost function admitting a zero-cost symbol, the rate per unit cost
    achieved by the proposed scheme is within a factor two of the
    information-theoretic optimum.
\end{abstract}

\section{Introduction}
\label{sec:intro}

Traditionally, data transmission in a communication system is based on
tight synchronization between the transmitter and the receiver. This
tight synchronization is usually achieved through either of two
strategies. In the first strategy, synchronization is achieved through
periodic transmission of pilot signals, followed by transmission of
information over the synchronized channel {\color{black} (see, e.g.,
\cite[Chapter 6.3]{proakis03})}. In the second strategy, data bits are
{\color{black} differentially encoded and then modulated}
({\color{black} e.g., differential pulse-position-modulation})
\cite[Chapter 4.3.2]{proakis03}, which implicitly achieves tight
synchronization.

The above strategies work well at high signal-to-noise ratios (SNRs) as
the energy overhead of achieving tight synchronization is negligible
compared to that of data transmission. However, in many applications,
such as wireless sensor networks, space communication, or in general any
communication system requiring high energy efficiency, communication by
necessity has to primarily take place in the low-SNR regime (due to the
concavity of the power-rate function). In such scenarios, the energy
overhead to achieve tight synchronization becomes significant and can
render the aforementioned strategies highly suboptimal in terms of
energy efficiency. In fact, requiring tight transmitter-receiver
synchronization can have arbitrarily large loss in performance in terms
of energy efficiency {\color{black} (see Example~\ref{ex:arbitrarily_bad}
in Section~\ref{sec:main})}.

\begin{figure}[htbp]
    \centering
    \includegraphics{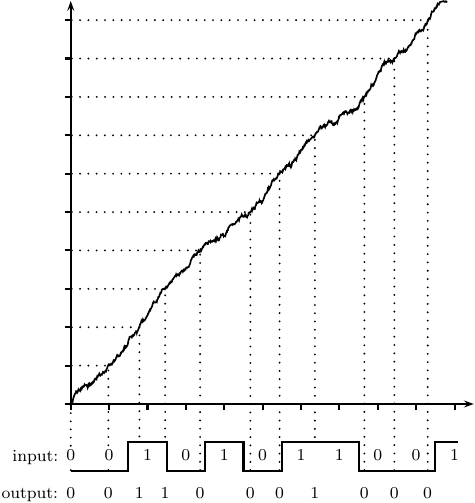}
    \caption{An example of unsynchronized transmitter-receiver clocks.
    The figure plots the value of the receiver clock ($y$-axis) as a
    function of the value of the reference clock at the transmitter
    ($x$-axis). The drift and jitter of the receiver clock are visible.
    For a transmitted input sequence, the lack of synchronization leads
    to duplications/{\allowbreak}deletions in the corresponding sampled
    output sequence at the receiver (illustrated here for the case
    without receiver noise).}
    \label{fig:motive}
\end{figure}

To mitigate this, in this paper, we develop and analyze a framework to
perform data transmission while only requiring loose synchronization
between the transmitter and the receiver.  To focus on the
energy-efficiency aspect, we choose the rate per unit cost (with energy
being a prime example of the cost) as our performance metric.  We model
synchronization errors through channel
duplications/{\allowbreak}deletions---an approach introduced
in~\cite{dobrushin67}. To motivate this model, consider a
transmitter-receiver pair with unsynchronized clocks, as illustrated in
Fig.~\ref{fig:motive}. Due to the absence of synchronization, the value
of the clock at the receiver exhibits drift and jitter with respect
to the value of the reference clock at the transmitter. This leads to
the receiver sampling the transmitted signal either faster than the
transmitter, leading to channel duplications, or slower, leading
to channel deletions.

Before we describe the contributions of this work in more detail in
Section~\ref{sec:intro_summary}, we provide a brief overview of related
work on energy-efficient communication and on channels with
synchronization errors.

\subsection{Related Work}
\label{sec:intro_related}

It is well known that the capacity per unit energy of a Gaussian channel
with noise variance $\eta^2$ is $1/(2\eta^2 \ln 2)$ and that this can
be achieved with appropriately designed pulse-position
modulation~\cite{golay49}. For a general discrete memoryless channel
(DMC), \cite{gallager88} has analyzed the reliability function of the
rate per unit cost. Subsequently, \cite{verdu90} has obtained a succinct
single-letter characterization for the capacity per unit cost of a DMC
with a general cost function. These results, however, strongly depend on
the channel being memoryless. As discussed next, synchronization errors
introduce memory in the channel, and thus the aforementioned results do
not apply.

The duplication/{\allowbreak}deletion/{\allowbreak}substitution channel
was introduced as a model for a channel with synchronization errors by
Dobrushin in 1967~\cite{dobrushin67}. Despite significant research
effort since then, the capacity of this channel is still not
known~\cite{diggavi06,mitzenmacher06,diggavi07,drinea07,kirsch10,fertonani10,
kanoria10,kalai10}. Indeed, even for one of the simplest versions of
this problem, the noiseless binary deletion channel, only loose bounds
on the capacity are known. For example, the recent
paper~\cite{mitzenmacher06} provides an approximation of the capacity of
the binary deletion channel to within a factor $9$ {\color{black} in
general}.  {\color{black} Tighter bounds have been obtained for some
specific regimes of the deletion probability: For instance, in
\cite{kanoria10} \cite{kalai10}, upper and lower bounds are provided and
vanishing gap is shown in the asymptotically small deletion-probability
regime; while improved upper bounds based on a numerical approach are
obtained in \cite{fertonani10}.} The main difficulty in analyzing these
channels arises from to the channel memory introduced by the duplications and
deletions, which prevents a direct application of the standard
information-theoretic tools.

It is worth emphasizing that the synchronization errors considered here
are those at the symbol level. There are other types of synchronization
issues. One such issue is frame synchronization, where errors are caused
by the incorrect identification of the location of the ``sync word'' in
the frame~\cite{massey72}. {\color{black} Thus, frame synchronization
deals with synchronization errors at the level of the block size as
compared to symbol-level synchronization considered here.}
Energy-efficient communication in the presence of such frame
asynchronism has been investigated in~\cite{chandar13}.

\subsection{Summary of Results}
\label{sec:intro_summary}

\begin{figure}[htbp]
    \centering
    \includegraphics{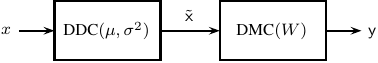}
    \caption{The end-to-end communication channel between an
    unsynchronized transmitter-receiver pair is modeled by concatenating
    a duplication/{\allowbreak}deletion channel $\DDC(\mu,\sigma^2)$ with
    a discrete memoryless channel $\DMC(W)$.}
    \label{fig:fig_IDC_DMC}
\end{figure}

In this paper, we consider communication channels which, in addition to
synchronization errors, exhibit receiver noise. We model the end-to-end
communication channel between the transmitter and the receiver as the
concatenation of two sub-channels, as illustrated in
Fig.~\ref{fig:fig_IDC_DMC}. The first sub-channel is a
duplication/{\allowbreak}deletion channel (DDC), which models
synchronization errors. This DDC is a sub-class of the
more general insertion/{\allowbreak}deletion/{\allowbreak}substitution
channels considered by Dobrushin \cite{dobrushin67} in that
\textit{random} insertions (as opposed to duplications only) are not
considered here. The second sub-channel is a noisy memoryless channel,
which models errors due to the receiver noise. The concatenation of the
two channels is a
duplication/{\allowbreak}deletion/{\allowbreak}substitution
channel.  The details of this model are discussed in
Section~\ref{sec:prob}.

We first study communication systems with synchronization errors
operating over Gaussian channels. We propose a new communication scheme
that requires only loose synchronization between the transmitter and the
receiver. {\color{black} Specifically, the scheme deals with the lack of
transmitter-receiver synchronization by incorporating guard spaces into
traditional pulse-position-modulation signaling.} Decoding at the
receiver is based on a sequence of independent hypothesis tests. When
the aforementioned durations are chosen appropriately, we show that the
scheme asymptotically achieves the information-theoretically optimal
performance in terms of the energy efficiency, i.e., the capacity per
unit energy.  In the process, we also establish that {\color{black} in
regimes of practical interest} the lack of transmitter-receiver
synchronization causes negligible loss in terms of energy efficiency.
{\color{black} We also consider the compound setting where even the
statistical properties of synchronization errors are not known precisely
a priori. For this setting, we show similar results by developing new
pulse-position-modulation waveforms, where the signal energy is spread
over increasing intervals and guard spaces of increasing lengths are
introduced.}

We then analyze communication systems with synchronization errors
operating over general DMCs and with a general cost function admitting a
zero-cost symbol. We generalize the proposed achievable scheme for the
Gaussian case to DMCs, and we show that the scheme achieves a rate per
unit cost within a factor two of the information-theoretic optimum.
Thus, while only loose bounds are known for the \emph{capacity} of the
general duplication/{\allowbreak}deletion/{\allowbreak}substitution
channel, we provide here a tight approximation for its \emph{capacity
per unit cost}. To establish this, we obtain an upper bound on the
capacity per unit cost of the channel in Fig.~\ref{fig:fig_IDC_DMC} by
considering the effect of the DDC as a specific way of encoding for the
DMC with an appropriately modified cost function. The upper bound is
then obtained by utilizing the characterization of the capacity per unit
cost for memoryless channels in~\cite{verdu90}.

\subsection{Organization}
\label{sec:intro_organization}

The remainder of the paper is organized as follows.
Section~\ref{sec:prob} provides the detailed description of the channel
model and the problem formulation.  The main results of the paper are
summarized in Section~\ref{sec:main}.  Section~\ref{sec:propose}
describes the proposed scheme for a general discrete memoryless channel
with synchronization errors.  Section~\ref{sec:converse} derives an
upper bound on its capacity per unit cost.
Section~\ref{sec:propose_gauss} discusses the proposed scheme for the
Gaussian channel with synchronization errors and establishes its
asymptotic optimality.  Lastly, Section~\ref{sec:propose_compound}
analyzes the performance of the scheme for the Gaussian channels with
synchronization errors where even the statistical properties of errors
are not known precisely a priori (i.e., the compound setting).

\section{Channel Model and Problem Statement}
\label{sec:prob}

We consider a duplication/{\allowbreak}deletion/{\allowbreak}substitution
channel with a cost constraint. The
duplication/{\allowbreak}deletion/{\allowbreak}substitution channel
consists of a DDC connected to a DMC as shown in
Fig.~\ref{fig:fig_IDC_DMC} in Section~\ref{sec:intro}.  The
duplication/{\allowbreak}deletion part of the channel models
synchronization errors (see Fig.~\ref{fig:motive} in
Section~\ref{sec:intro}), the substitution part models noise.

The DDC maps the input sequence $(x[1], \dots, x[T])\in\mc{X}^T$ to the
output sequence $(\tilde{\msf{x}}[1], \dots,
\tilde{\msf{x}}[\msf{L}])\in\mc{X}^\msf{L}$ for some random length
$\msf{L}$, where here and in the following we use sans-serif font to
denote random variables. The actions of the DDC are governed by the \iid
sequence of states $(\msf{s}[1], \dots,
\msf{s}[T])\in\{0,1,2,\dots\}^T$.  State $s[t]$ describes how many times
input symbol $x[t]$ appears at the output of the DDC.

Formally, the total number of output bits is given by
\begin{equation*}
    \msf{L} \defeq \sum_{t=1}^T \msf{s}[t].
\end{equation*}
Observe that $\msf{L}$ is a random variable depending on the state
sequence of the DDC. Define for each $\ell\in \{1, 2, \dots, \msf{L}\}$
the random variable
\begin{equation*}
    \msf{t}[\ell] \defeq \min\biggl\{ \tau : \sum_{j=1}^{\tau} \msf{s}[j] \geq \ell \biggr\}.
\end{equation*}
The relationship between the input and output of the DDC is then given
by
\begin{equation*}
    \tilde{\msf{x}}[\ell] \defeq x[\msf{t}[\ell]].
\end{equation*}
We illustrate the operation of the DDC with an example.
\begin{example}
    \begin{equation*}
        \begin{array}{ccccccccccc}
            \bm{x} & = & x[1] & x[2] & x[3] & x[4] & x[5] & x[6] & \\
            \bm{\msf{s}} & = & 1 & 1 & 2 & 1 & 0 & 2 & \\
            \bm{\msf{t}} & = & 1 & 2 & 3 & 3 & 4 & 6 & 6  \\
            \bm{\tilde{\msf{x}}} & = & x[1] & x[2] & x[3] & x[3] & x[4] & x[6] & x[6]
        \end{array}
    \end{equation*}
    In the example, $T=6$ and $\msf{L}=7$. Here $\bm{x}$ is the vector of
    inputs and $\bm{\tilde{\msf{x}}}$ is the vector of outputs of the
    DDC. $\bm{\msf{s}}$ is the vector of states and the
    corresponding vector of sampling times is $\bm{\msf{t}}$. For
    concreteness, assume
    \begin{align*}
        \bm{x} & =
        \begin{pmatrix}
            0 & 0 & 1 & 0 & 1 & 0
        \end{pmatrix}. \\
        \shortintertext{Then the output of the DDC is} \\
        \tilde{\bm{\msf{x}}} & =
        \begin{pmatrix}
            0 & 0 & 1 & 1 & 0 & 0 & 0
        \end{pmatrix},
    \end{align*}
    see also Fig.~\ref{fig:motive} in Section~\ref{sec:intro}.
\end{example}

We denote by
\begin{align*}
    \mu & \defeq \E(\msf{s}[1]) \\
    \shortintertext{and}\\
    \sigma^2 & \defeq \var(\msf{s}[1])
\end{align*}
the mean and variance of the
duplication/{\allowbreak}deletion process,
respectively, and we refer to any DDC with those parameters as
$\DDC(\mu, \sigma^2)$.  Here, $\mu$ and $\sigma^2$ can be interpreted as
capturing the drift and jitter of the receiver clock, respectively. In
most situations arising in practice, the parameter $\mu$ is close to
$1$: {\color{black} For instance, in 3GPP-Long Term Evolution (LTE), one
of the modes (LTE-TDD) requires clock accuracy of $50$ parts per billion
\cite{bladsjo13}, i.e., $\mu=1 \pm 5\cdot 10^{-8}$; even in a popular
compact wireless sensor node, Mica2 mote, the clock accuracy is
specified as $40$ parts per million \cite{gankumsri03}, i.e., $\mu = 1
\pm 4\cdot 10^{-5}$.}

Our results will be presented for arbitrary
duplication/{\allowbreak}deletion processes with finite mean and variance.
For illustrative purposes, we present a commonly used special case of
this setting.
\begin{example}
    \label{eg:deletion}
    One commonly used definition of the state process is
    \begin{equation*}
        \msf{s}[t] =
        \begin{cases}
            1, & \text{w.p. } 1-d, \\
            0, & \text{w.p. } d,
        \end{cases}
    \end{equation*}
    for some parameter $d$. This results in the so-called \emph{deletion
    channel}, which deletes each input symbol independently with
    probability $d$.
\end{example}

The output of the $\DDC(\mu,\sigma^2)$ is then fed into a discrete
memoryless channel $\DMC(W)$ described by the distribution
$W(\cdot|\tilde{x})$ of the channel output $\msf{y}\in\mc{Y}$ given the
channel input $\tilde{x}\in\mc{X}$. The
duplication/{\allowbreak}deletion/{\allowbreak}substitution channel is the
(random) mapping from $x$ to $\msf{y}$ described by the concatenation of
the $\DDC(\mu,\sigma^2)$ and the $\DMC(W)$.

{\color{black}
\begin{remark}
    This model is motivated by a widely-used digital demodulation
    architecture consisting of an envelope detector followed by matched
    filters (see for example \cite[pages 302--304]{proakis03}). The
    model captures the duplication/deletion of symbols in such an
    architecture due to the lack of transmitter-receiver
    synchronization. However, this model does not account for the fact
    that over or under sampling of the received signal introduces
    different signal attenuations. As a consequence, the model as
    defined here has physical meaning only when $\mu\approx 1$, which is
    exactly the scenario motivating the analysis in this paper. For
    other regimes of $\mu$, the model may need to be suitably modified
    (for example by normalizing the channel output) to account for the
    signal attenuation due to over or under sampling.
\end{remark}
}

The goal is to maximize the number of bits reliably transmitted per unit
cost over this duplication/{\allowbreak}deletion/{\allowbreak}substitution
channel formed by the concatenation of the $\DDC(\mu,\sigma^2)$ with the
$\DMC(W)$. We adopt the framework of~\cite{gallager88,verdu90}. The
\emph{cost function} $c\from\mc{X}\to\Rp$ associates with each input
symbol $x\in\mc{X}$ the cost $c(x)$ incurred by transmitting $x$ over
the channel. We make the assumption that $\mc{X}$ contains a free
input symbol, and without loss of generality we label this symbol as
$0$. In other words, $0\in\mc{X}$ and $c(0)=0$. For an input sequence
$(x[1], \dots, x[T])\in\mc{X}^T$, the cost is given by
\begin{equation*}
    c\bigl( (x[1], \dots, x[T]) \bigr)
    \defeq \sum_{t=1}^T c(x[t]).
\end{equation*}
A \emph{$(T,M,P,\varepsilon)$ code} consists of an
encoder-decoder pair $f^T$ and $\{\varphi^\ell\}_{\ell=0}^{\infty}$ and $M$
codewords. The encoder $f^T$ maps the message $m\in\{1,2,\ldots,M\}$ to
a codeword
\begin{equation*}
    (x_{m}[1], \dots ,x_{m}[T]), %\quad m\in\{1,2,\ldots,M\}
\end{equation*}
of length $T$ and cost at most $P$. The decoder consists of several
sub-decoders $\{\varphi^\ell\}_{\ell=0}^{\infty}$, one for each possible
realization $\ell$ of $\msf{L}$, and has average (assuming equiprobable
messages) probability of decoding error at most $\varepsilon$.

\begin{definition}
    A rate $\hat{R}$ per unit cost is \emph{achievable} if for every
    $\varepsilon>0$ and every large enough $M$ there exists a
    $(T,M,P,\varepsilon)$ code satisfying\footnote{Throughout this
        paper, $\log(\cdot)$ and $\ln(\cdot)$ denote the logarithms to
    the base $2$ and $e$, respectively.} $\log(M)/P\geq \hat{R}$.  The
    \emph{capacity per unit cost} $\hat{C}$ is the supremum of
    achievable rates per unit cost.\footnote{{\color{black} In general,
    $\hat{C}$ might take the value of $-\infty$ in case that the set of
    achievable rates is empty. Hence, $\hat{C}$ is well defined on the
    extended real line $\bar{\R}$. For the case of a zero-cost input
    symbol as considered here, we will see later that $\hat{C} \geq 0$.}}
\end{definition}

It can be seen from the above definition that, since the receiver knows
the realization of $\msf{L}$ (i.e., it knows the random length of the
transmission), the present paper considers the scenario of a one-shot
transmission.

Throughout this paper, we are interested in $\hat{C}(\mu,\sigma^2,W)$,
the capacity per unit cost of the
duplication/{\allowbreak}deletion/{\allowbreak}substitution
channel given by the $\DDC(\mu,\sigma^2)$ concatenated with the
$\DMC(W)$. We also consider the compound capacity per unit cost
$\hat{C}([\mu_1,\mu_2],[\sigma_1^2,\sigma_2^2],W)$, for which the
encoder and decoder have to be able to operate on any $\DDC(\mu,\sigma)$
with $\mu\in[\mu_1,\mu_2]$ and $\sigma\in[\sigma_1^2,\sigma_2^2]$
without knowledge of the actual values of $\mu$ and $\sigma^2$. This
compound setting is of practical relevance, since usually the mean clock
drift $\mu$ is only specified as an interval (and might indeed be slowly
time varying) and is hence not known exactly at the transmitter or the
receiver. We also treat the Gaussian version of the problem, where the
output of the $\DDC(\mu,\sigma^2)$ is subject to additive Gaussian noise
of mean zero and variance $\eta^2$. The cost function is in this case
the signal energy, i.e., $c(x)=x^2$. With slight abuse of notation, we
refer to the capacity per unit energy in this case as
$\hat{C}(\mu,\sigma^2,\mc{N}(0,\eta^2))$.

\section{Main Results}
\label{sec:main}

In this section, we summarize the main results; their proofs are
discussed in subsequent sections. We start with the results for Gaussian
channels with synchronization errors for the case where the statistics
$\mu$ and $\sigma^2$ of the duplication/{\allowbreak}deletion channel are
known at the transmitter and the receiver.

\begin{theorem}
    \label{thm:Chat_Gauss}
    The Gaussian channel with synchronization errors having
    duplication/{\allowbreak}deletion process with mean $\mu$ and variance
    $\sigma^2$ and having noise power $\eta^2$ has capacity per unit
    energy
    \begin{equation*}
        \hat{C}(\mu, \sigma^2, {\cal N}(0,\eta^2))
        = \frac{\mu}{2\eta^2\ln 2}.
    \end{equation*}
\end{theorem}

Recall that the capacity per unit energy of the Gaussian channel is
$1/(2\eta^2\ln 2)$. Furthermore, as discussed in Section~\ref{sec:prob},
the mean $\mu$ of the duplication/{\allowbreak}deletion process is
typically close to $1$. Hence, Theorem~\ref{thm:Chat_Gauss} implies
that the lack of synchronization results in only negligible loss in the
capacity per unit energy.

To establish achievability, we propose a new communication scheme that
jointly performs data modulation and loose synchronization. To this end,
we develop signaling waveforms where the signal energy is spread over
increasing intervals and guard spaces of increasing lengths are
introduced. Decoding at the receiver is based on a sequence of
independent hypothesis tests, which are carefully chosen to account for
the uncertainty arising due to the lack of tight synchronization. In
Section~\ref{sec:propose_gauss}, we show that, by appropriately choosing
the aforementioned durations, the probability of error can be made
arbitrarily small for any rate per unit energy up to $\hat{C}$. The
details of the analysis of this scheme are reported in
Section~\ref{sec:propose_gauss}. The upper bound in
Theorem~\ref{thm:Chat_Gauss} follows as a special case of the upper
bound derived for a general DMC with synchronization errors, discussed
in Theorem~\ref{thm:Chat_DMC} below.

\begin{remark}
    From the above results, it follows that the capacity per unit cost
    $\hat{C}(\mu,\sigma^2,W)$ depends on the distribution of the
    duplication/{\allowbreak}deletion process $\msf{s}$ only through its
    mean.  This is because $\sigma^2$ is finite and we have a zero-cost
    symbol.  Hence, one can use a long block of the zero-cost symbol to
    smooth the variation of the duplication/deletion process. 
\end{remark}

Next, consider the case where the exact statistical properties of the
duplication/{\allowbreak}deletion process are not known a priori. Instead,
we only know a range for each parameter, i.e., the mean $\mu$ is in
$[\mu_1, \mu_2]$ and the variance in $[0, \sigma^2]$. We are interested
in a communication scheme that works simultaneously for every possible
set of parameters in this range. As pointed out in
Section~\ref{sec:prob}, this compound setting is of practical relevance,
since the precision of the transmitter and receiver clocks are usually
only known to lie within some range.

\begin{theorem}
    \label{thm:Chat_compound}
    The class of Gaussian channels with synchronization errors having
    duplication/{\allowbreak}deletion process with mean
    $\mu\in[\mu_1,\mu_2]$ and variance upper bounded by $\sigma^2$ and
    having noise power $\eta^2$ has compound capacity per unit
    energy
    \begin{equation*}
        \hat{C} ([\mu_1,\mu_2], [0, \sigma^2], {\cal N}(0,\eta^2))
        = \frac{\mu_1}{2\eta^2\ln 2}.
    \end{equation*}
\end{theorem}

Comparing Theorems~\ref{thm:Chat_Gauss}
and~\ref{thm:Chat_compound}, we see that
\begin{equation*}
        \hat{C} ([\mu_1,\mu_2], [0, \sigma^2], {\cal N}(0,\eta^2))
        = \min_{\mu\in[\mu_1,\mu_2]}\hat{C} (\mu, \sigma^2, {\cal N}(0,\eta^2)).
\end{equation*}
Since a scheme for the compound setting must work for any possible value
of $\mu$ and $\sigma^2$ of the duplication/{\allowbreak}deletion process,
it is clear that it must work for the worst one, so that
\begin{align*}
        \hat{C} ([\mu_1,\mu_2], [0, \sigma^2], {\cal N}(0,\eta^2))
        \leq \min_{\mu\in[\mu_1,\mu_2]}\hat{C} (\mu, \sigma^2, {\cal N}(0,\eta^2)) 
        = \frac{\mu_1}{2\eta^2\ln 2}.
\end{align*}
Theorem~\ref{thm:Chat_compound} thus shows that there is no further
loss beyond this resulting from the lack of precise knowledge of the
duplication/{\allowbreak}deletion statistics at the transmitter and the
receiver.  The proof of Theorem~\ref{thm:Chat_compound} is presented
in Section~\ref{sec:propose_compound}.

Finally, consider a duplication/{\allowbreak}deletion channel
$\DDC(\mu,\sigma^2)$ concatenated with a general discrete memoryless
channel $\DMC(W)$ specified by its transition probability matrix
$W\from\mc{X}\to\mc{Y}$. Furthermore, consider an arbitrary cost function
$c\from\mc{X}\to\Rp$. As mentioned in Section~\ref{sec:prob}, we assume
that $0\in\mc{X}$ and $c(0)=0$.  Then, the following bounds hold on the
capacity per unit cost.

\begin{theorem}
    \label{thm:Chat_DMC}
    The duplication/{\allowbreak}deletion/{\allowbreak}substitution
    channel consisting of a $\DDC(\mu,\sigma^2)$ concatenated with
    a $\DMC(W)$ has capacity per unit cost $\hat{C}(\mu,\sigma^2, W)$
    satisfying
    \begin{align*}
        \frac{\mu}{2} \sup_{x\in\mc{X}\setminus\{0\}}
        \frac{D(W(\cdot|x) \Vert W(\cdot|0))}{c(x)} 
        \leq \hat{C}(\mu,\sigma^2, W)
        \leq
        \mu \sup_{x\in\mc{X}\setminus\{0\}}
        \frac{D(W(\cdot|x) \Vert W(\cdot|0))}{c(x)}
    \end{align*}
    where $D(P \Vert Q)$ is the Kullback-Leibler divergence between
    distributions $P$ and $Q$ and where $0\in\mc{X}$ is an input symbol
    with zero cost.
\end{theorem}

Theorem~\ref{thm:Chat_DMC} approximates the capacity per unit cost of a
general DMC with synchronization errors and general cost function
admitting a zero-cost symbol to within a factor two. In contrast, recall
from Section~\ref{sec:intro} that for the \emph{capacity}, even of the
noiseless deletion channel, only loose bounds are known despite over
four decades since the introduction of the model in~\cite{dobrushin67}.

It was shown in~\cite{verdu90} that the capacity per unit cost $\hat{C}(W)$
of a $\DMC(W)$ with zero-cost symbol $0$ is
\begin{equation}
    \label{eq:verdu}
    \hat{C}(W)
    = \sup_{x\in\mc{X}\setminus\{0\}}
    \frac{D(W(\cdot|x) \Vert W(\cdot|0))}{c(x)}.
\end{equation}
Thus, from the lower and upper bounds in Theorem~\ref{thm:Chat_DMC}, we
obtain the following corollary, showing that the loss due to
synchronization errors is within a factor between $\mu/2$ and $\mu$.
\begin{corollary}
    \label{cor:Chat_DMC}
    The duplication/{\allowbreak}deletion/{\allowbreak}substitution
    channel consisting of a $\DDC(\mu,\sigma^2)$ concatenated with
    a $\DMC(W)$ has capacity per unit cost $\hat{C}(\mu,\sigma^2, W)$
    satisfying
    \begin{equation*}
        \frac{\mu}{2} \hat{C}(W)
        \leq \hat{C}(\mu,\sigma^2, W)
        \leq \mu \hat{C}(W).
    \end{equation*}
\end{corollary}
{\color{black}
\begin{remark}
    The upper bound in Corollary~\ref{cor:Chat_DMC} in fact holds irrespective of whether
    there is a zero-cost symbol or not (as will become clear from the proof in Section~\ref{sec:converse}).
\end{remark}
}

The achievability in Theorem~\ref{thm:Chat_DMC} is established by
generalizing the proposed scheme for the Gaussian channel with
synchronization errors to DMCs. The details are discussed in
Section~\ref{sec:propose}. For the upper bound on the capacity per unit
cost in Theorem~\ref{thm:Chat_DMC}, we treat the effect of the DDC as a
specific way of encoding for the DMC with an appropriately modified cost
function. The upper bound is then obtained by utilizing the
characterization of the capacity per unit cost for {\em memoryless}
channels in \cite{verdu90}. Section~\ref{sec:converse} provides the
details.

We conclude this section by illustrating through an example that the
conventional schemes based on tight synchronization between the
transmitter and the receiver can be highly suboptimal in terms of their
rate per unit cost.

\begin{example}\label{ex:arbitrarily_bad}
    Let us consider the simplest synchronized communication setting: a
    channel with binary input and no noise, i.e., $W(x|x)=1$ for
    $x\in\{0,1\}$.  Further, let the cost function be the number of ones
    transmitted, i.e., $c(x)=x$. This is a DMC, and by
    \eqref{eq:verdu}, its capacity per unit cost is
    \begin{equation*}
        \hat{C}(W)
        = \frac{D(W(\cdot|1) \Vert W(\cdot|0))}{c(1)}
        = \infty.
    \end{equation*}

    Let us now consider the scenario where the transmitter and the
    receiver are no longer perfectly synchronized. Specifically, the
    input signals are first corrupted by a deletion channel with
    deletion probability $d\in(0,1)$ (see Example~\ref{eg:deletion} in
    Section~\ref{sec:prob} for a formal definition of this special case
    of a DDC), before being sent over the aforementioned noiseless
    channel $W$.

    Consider first the operation of conventional schemes based on tight
    synchronization. In this example, we take this to mean any scheme
    that detects and corrects deletions without letting them accumulate.
    This definition applies to schemes using pilots {\color{black}
    \cite[Chapter 6.3]{proakis03}} as well as to schemes using
    differential modulation {\color{black} \cite[Chapter
    4.3.2]{proakis03}}. To maintain tight synchronization, the channel
    inputs cannot contain too many consecutive zeros (since otherwise
    deletions would accumulate without any way of correcting for them).
    On average, we expect to see about one deleted bit every $1/d$
    transmitted bits. Thus, roughly every $1/d$ channel inputs needs to
    be a $1$ at a cost of $c(1)=1$. Now, over a block of $1/d$ binary
    channel uses, we can reliably transmit at most $1/d$ bits. Hence,
    the rate per unit cost achieved by any scheme based on tight
    synchronization is at most
    \begin{equation*}
        \hat{R}_{\text{sync}}(d,W) \leq \frac{1}{d} < \infty.
    \end{equation*}

    On the other hand, from Corollary~\ref{cor:Chat_DMC}, the
    communication scheme proposed in this paper achieves a rate per unit
    cost that is within a factor of $\mu/2=(1-d)/2$ of the capacity per
    unit cost $\hat{C}(W)$ of the underlying $\DMC(W)$. Hence, the
    capacity per unit cost with synchronization errors is
    \begin{equation*}
        \hat{C}(d,W) \geq \frac{1-d}{2}\hat{C}(W) = \infty.
    \end{equation*}
    Thus, even in the presence of synchronization errors, the rate per
    unit cost achieved by the proposed scheme is arbitrarily large. This
    illustrates that the improvement in the rate per unit cost achieved by the
    proposed scheme over schemes based on tight synchronization can be
    unbounded.
\end{example}

\section{Proof of Lower Bound in Theorem~\ref{thm:Chat_DMC}}
\label{sec:propose}

In this section, we propose a coding scheme that achieves the lower
bound on the rate per unit cost stated in Theorem~\ref{thm:Chat_DMC} for
a general DMC with synchronization errors. The scheme uses a type of
pulse-position modulation. To send message $m$, the encoder sends a
burst of symbols $x^\star\in\mc{X}\setminus\{0\}$ at a position
corresponding to this message. The decoder searches for the location of
the pulse using a sliding window. Once the pulse is located, the decoder
checks which decision region it is in and declares the corresponding
message. In order to deal with duplications and deletions, guard spaces
need to be introduced around the pulses and the decision regions need to
be chosen judiciously. We proceed with a detailed description of the
scheme and its analysis.

\emph{Encoding:} Fix a target error probability $\varepsilon\in(0,1)$
and a number $\delta\in(0,1)$. Let $x^\star\in\mc{X}\setminus\{0\}$ be a fixed
nonzero channel input. The codeword for message $m\in\{1,2,\ldots,M\}$
is
\begin{equation*}
    \bm{x}_m
    \defeq \bigl( \bm{0}_{(m-1)N}, x^\star\cdot \bm{1}_B, \bm{0}_{N-B}, \bm{0}_{(M-m)N} \bigr), \\
\end{equation*}
where
\begin{align*}
    N & \defeq \big\lceil 36M\sigma^2/(\mu^2\varepsilon) \big\rceil
    = \Theta(M), \\
    \intertext{and}
    B & \defeq \bigg\lfloor \frac{(2+\delta)\log(M)}{\mu D(W(\cdot|x^\star)\Vert W(\cdot|0))} \bigg\rfloor
    = \Theta(\log(M)).
\end{align*}
Thus, to communicate message $m$, the transmitter sends a pulse of
symbols $x^\star$ at position $(m-1)N+1$ and of duration $B$. Observe that
between adjacent possible pulse positions is a guard space of
$N-B$ zeroes. The block length of this code is $T=MN$ and the cost
of each codeword is
\begin{equation}
    \label{eq:power}
    P = B c(x^\star).
\end{equation}

\emph{Decoding:} Recall that the output $\bm{\msf{y}}$ of the channel
has length $\msf{L}$. The decoder forms the subsequences
\begin{equation*}
    \bm{\msf{y}}_\ell
    \defeq \bigl( \msf{y}[\ell], \msf{y}[\ell+1], \ldots, \msf{y}[\ell+\floor{B\mu-\beta}-1] \bigr)
\end{equation*}
of length $\floor{B\mu-\beta}$ for $\ell\in \{ 1, 2, \dots,
\msf{L}-\floor{B\mu-\beta}+1\}$ with
\begin{equation*}
    \beta \defeq \sqrt{4B\sigma^2/\varepsilon}
    = \Theta(\log^{1/2}(M)).
\end{equation*}
Similarly, define the subsequences
\begin{equation*}
    \tilde{\bm{\msf{x}}}_\ell
    \defeq \bigl( \tilde{\msf{x}}[\ell], \tilde{\msf{x}}[\ell+1],
    \ldots, \tilde{\msf{x}}[\ell+\floor{B\mu-\beta}-1] \bigr)
\end{equation*}
of the output of the DDC / input to the DMC (not observable at the
receiver). Finally, define the open intervals $\tilde{\mc{D}}_m \defeq
((m-1)N\mu+1-\nu, (m-1)N\mu+1+\nu)$ for $m\in\{2,\dots,M\}$ and define
the decision regions
\begin{equation*}
    \mc{D}_m \defeq \\
    \begin{cases}
        \{1\},  & \text{for $m=1$} \\
        \N\cap\tilde{\mc{D}}_m,
        & \text{for $m\in\{2,\dots,M\}$}
    \end{cases}
\end{equation*}
with
\begin{equation*}
    \nu \defeq \sqrt{4MN\sigma^2/\varepsilon} = \Theta(M).
\end{equation*}
In words, the decision region $\mc{D}_m$ for message $m$ consists of all
integer points within distance $\nu$ of $(m-1)N\mu+1$.

The receiver performs independent hypothesis tests for each $\bm{\msf{y}}_\ell$
for the two hypotheses
\begin{align*}
    H^0 & \defeq \{ \tilde{\bm{\msf{x}}}_\ell = \bm{0} \}, \\
    H^1 & \defeq \{ \tilde{\bm{\msf{x}}}_\ell = x^\star\cdot\bm{1} \}.
\end{align*}
Let $\hat{\msf{H}}_\ell$ be the decision of the hypothesis test for
$\bm{\msf{y}}_\ell$. The receiver declares that message $m$ was sent if
$\hat{\msf{H}}_\ell = H^1$ for \emph{some} $\ell\in\mc{D}_m$ and
$\hat{\msf{H}}_\ell = H^0$ for \emph{all} $\ell\in\mc{D}_{m'}$ with $m'\neq m$.
If no such $m$ exists, an error is declared.

In order for the decoder to be well defined, we need to ensure that the
decision regions are disjoint, i.e., that $\mc{D}_m \cap \mc{D}_{m'}
=\emptyset$ for $m\neq m'$. This is the case since, by the definitions of
$N$ and $\nu$,
\begin{align}
    \label{eq:disjoint}
    N\mu
    & \geq \sqrt{N}\sqrt{\frac{36M\sigma^2}{\mu^2\varepsilon}}\mu \nonumber\\
    & = 3\sqrt{\frac{4MN\sigma^2}{\varepsilon}} \nonumber\\
    & = 3\nu,
\end{align}
so that
\begin{equation*}
    mN\mu+1-\nu \geq (m-1)N\mu+1+\nu
\end{equation*}
for all $m$.

\emph{Error Analysis:} Assume that message $m$ was sent. Let
$\mc{E}_{1,\ell}$ be the event that $\hat{\msf{H}}_\ell = H^0$, and
$\mc{E}_{2,\ell}$ be the event that $\hat{\msf{H}}_\ell = H^1$. Define
the missed-detection event
\begin{align*}
    \mc{E}_1 & \defeq \bigcap_{\ell\in\mc{D}_m}\mc{E}_{1,\ell} \\
    \intertext{and the false-alarm event}
    \mc{E}_2 & \defeq \bigcup_{m'\neq m}\bigcup_{\ell\in\mc{D}_{m'}}\mc{E}_{2,\ell}.
\end{align*}
The probability of decoding error for message $m$ is equal to
$\Pp_m(\mc{E}_1\cup\mc{E}_2)$, where $\Pp_m$ denotes probability
conditioned on message $m$ being sent.

We continue by upper bounding this probability. It will be convenient to
define two auxiliary events isolating the behavior of the DDC.  Let
$\mc{E}_3$ be the event that the total number of symbols in
$\tilde{\bm{\msf{x}}}$ resulting from the first $(m-1)N$ transmitted
symbols is outside the interval $((m-1)N\mu-\nu,(m-1)N\mu+\nu)$, and let
$\mc{E}_4$ be the event that the number of symbols in
$\tilde{\bm{\msf{x}}}$ resulting from symbols transmitted during time
slots $(m-1)N+1$ to $(m-1)N+B$ is outside the interval
$(B\mu-\beta,B\mu+\beta)$. We have
\begin{align}
    \label{eqn:pe_dmc}
    \Pp_m(\mc{E}_1\cup\mc{E}_2) 
    &= \Pp_m(\mc{E}_1\cup\mc{E}_2\mid\mc{E}_3\cup\mc{E}_4)\Pp_m(\mc{E}_3\cup\mc{E}_4) 
    +\Pp_m(\mc{E}_1\cup\mc{E}_2\mid\mc{E}_3^c\cap\mc{E}_4^c)\Pp_m(\mc{E}_3^c\cap\mc{E}_4^c) \nonumber\\
    &\leq \Pp_m(\mc{E}_3\cup\mc{E}_4)
    + \Pp_m(\mc{E}_1\cup\mc{E}_2\mid\mc{E}_3^c\cap\mc{E}_4^c) \nonumber\\
    &\leq \Pp_m(\mc{E}_3) + \Pp_m(\mc{E}_4)
    + \Pp_m(\mc{E}_1\mid\mc{E}_3^c \cap \mc{E}_4^c) 
    + \Pp_m(\mc{E}_2\mid\mc{E}_3^c \cap \mc{E}_4^c).
\end{align}
The first two probabilities correspond to the events that the
$\DDC(\mu,\sigma^2)$ is not well behaved and the last two correspond to
the two possible detection errors caused by the $\DMC(W)$ conditioned on
the behavior of the DDC to be as expected.

We continue by upper bounding each of the terms in \eqref{eqn:pe_dmc} in
turn. By Chebyshev's inequality,
\begin{align}
    \label{eq:e1}
    \Pp_m(\mc{E}_3)
    & = \Pp_m\bigl( \big\lvert {\textstyle\sum_{t=1}^{(m-1)N}} \msf{s}[t] - (m-1)N\mu \big\rvert
    \geq \nu \bigr) \nonumber \nonumber\\
    & \leq \frac{(m-1)N\sigma^2}{\nu^2} \nonumber \\
    & \leq \varepsilon/4, \\
    \intertext{and}
    \label{eq:e2}
    \Pp_m(\mc{E}_4)
    & = \Pp_m\bigl( \big\lvert {\textstyle\sum_{t=(m-1)N+1}^{(m-1)N+B}} \msf{s}[t] - B\mu \big\rvert
    \geq \beta \bigr) \nonumber \\
    & \leq \frac{B\sigma^2}{\beta^2} \nonumber \\
    & = \varepsilon/4,
\end{align}
where we have used the definitions of $\nu$ and $\beta$, respectively.

We proceed with the analysis of $\Pp_m(\mc{E}_1|\mc{E}_3^c \cap
\mc{E}_4^c)$ and $\Pp_m(\mc{E}_2|\mc{E}_3^c \cap \mc{E}_4^c)$. The
following is the key observation for this analysis. Conditioned on
message $m$ being sent and on $\mc{E}_3^c\cap\mc{E}_4^c$, the elements
in the decision regions satisfy the following two properties for
$M$ large enough (not depending on $m$):
\begin{enumerate}
    \item For every $\ell\in\mc{D}_{m'}$ with $m'\neq m$, we have
        $\tilde{\bm{\msf{x}}}_\ell=\bm{0}$. Hence, the
        symbols in the subsequence $\bm{\msf{y}}_\ell$ are
        \iid with distribution $W(\cdot|0)$.
    \item There exists at least one $\ell\in\mc{D}_m$ such that
        $\tilde{\bm{\msf{x}}}_\ell=x^\star\cdot\bm{1}$. Hence, the
        symbols in the subsequence $\bm{\msf{y}}_\ell$
        are \iid with distribution  $W(\cdot|x^\star)$.
\end{enumerate}

We start by proving property~1. By construction of the codewords, and
since the DDC part of the channel can only delete and duplicate symbols
but never ``create'' them, we only need to argue that the burst of
symbol $x^\star$ sent in block $m$ by the transmitter cannot be shifted
into the decision region $\mc{D}_{m'}$.

Assume first $m'< m$. The right-most element of $\mc{D}_{m'}$ is at
position less than or equal to $(m-2)N\mu+1+\nu$, and therefore the
right-most element of $\tilde{\bm{\msf{x}}}_\ell$ with
$\ell\in\mc{D}_{m'}$ is at position at most
\begin{equation*}
    (m-2)N\mu+\nu+B\mu.
\end{equation*}
Now, conditioned on $\mc{E}_3^c$, there are at least
$(m-1)N\mu-\nu$ symbols $0$ in $\tilde{\bm{\msf{x}}}$ before the
first symbol $x^\star$. For there to be no overlap, it is sufficient to
argue that
\begin{equation*}
    (m-2)N\mu+\nu+B\mu \leq (m-1)N\mu-\nu,
\end{equation*}
or, equivalently, that
\begin{equation*}
    N\mu \geq 2\nu+B\mu.
\end{equation*}
This holds for $M$ large enough since we have $N\mu \geq 3\nu$
by~\eqref{eq:disjoint}, and since $\nu=\Theta(M)$ whereas $B =
\Theta(\log(M))$.

Assume then that $m' > m$. The left-most element of any
$\tilde{\bm{\msf{x}}}_\ell$ with $\ell\in\mc{D}_{m'}$ is at position at
least $mN\mu+1-\nu$. Conditioned on $\mc{E}_3^c$, there are at
most $(m-1)N\mu+\nu$ symbols $0$ before the first symbol
$x^\star$ in $\tilde{\bm{\msf{x}}}$. Conditioned on $\mc{E}_4^c$, the
burst of symbol $x^\star$ in $\tilde{\bm{\msf{x}}}$ is of length at most
$B\mu+\beta$. For there to be no overlap, it is sufficient to
argue that
\begin{equation*}
    (m-1)N\mu+\nu+B\mu+\beta < mN\mu+1-\nu,
\end{equation*}
or, equivalently, that
\begin{equation*}
    N\mu \geq 2\nu+B\mu+\beta.
\end{equation*}
This holds for $M$ large enough by the same argument as in the last
paragraph since $\beta=\Theta(\log^{1/2}(M))$. Together, this proves
property~1.

To prove property~2, observe that, conditioned on $\mc{E}_4^c$,
the burst of symbols $x^\star$ in $\tilde{\bm{\msf{x}}}$ is of length at
least $B\mu-\beta$. Further, conditioned on $\mc{E}_3^c$, this
burst must start at the receiver in the interval
\begin{equation*}
    \N\cap( (m-1)N\mu+1 - \nu, (m-1)N\mu+1 + \nu ) = \mc{D}_m.
\end{equation*}
Since the subsequences $\tilde{\bm{\msf{x}}}_\ell$ have length $\floor{B\mu -
\beta}$, these two statements show that there exists at least one
$\ell\in\mc{D}_m$ such that $\tilde{\bm{\msf{x}}}_\ell=x^\star\cdot\bm{1}$.

The two properties allow us to analyze the events $\mc{E}_{1,\ell}$ and
$\mc{E}_{2,\ell}$. Recall that the hypothesis test on
$\bm{\msf{y}}_\ell$ is performed under the assumption that either
$\tilde{\bm{\msf{x}}}_\ell=\bm{0}$ or
$\tilde{\bm{\msf{x}}}_\ell=x^\star\cdot\bm{1}$. Fix a threshold for the hypothesis
test of $\bm{\msf{y}}_\ell$ such that the probability of missed
detection satisfies
\begin{equation}
    \label{eq:H0}
    \Pp(\msf{H}_\ell=H^0\mid\tilde{\bm{\msf{x}}}_\ell=x^\star\cdot\bm{1})
    \leq \varepsilon/4.
\end{equation}
By Stein's lemma (see, e.g., \cite[Theorem 12.8.1]{cover91}), we then have that
the probability of false alarm of the optimal test is upper bounded by
\begin{align}
    \label{eq:H1}
    \Pp(\msf{H}_\ell=H^1\mid\tilde{\bm{\msf{x}}}_\ell=\bm{0})
    & \leq 2^{-\floor{B\mu-\beta}D(W(\cdot|x^\star)||W(\cdot|0)) + o(B\mu-\beta)} \nonumber\\
    & = 2^{-B\mu D(W(\cdot|x^\star)||W(\cdot|0)) + o(\log(M))}
\end{align}
as $M\to\infty$, and where we have used that $B=\Theta(\log(M))$ and
$\beta=\Theta(\log^{1/2}(M))$.

Consider then the value of $\ell\in\mc{D}_m$ guaranteed by
property~2. For this $\ell$, we have by \eqref{eq:H0},
\begin{align}
    \label{eq:e3}
    \Pp_m(\mc{E}_1\mid\mc{E}_3^c\cap\mc{E}_4^c)
    & \leq \Pp_m(\mc{E}_{1,\ell}\mid\mc{E}_3^c\cap\mc{E}_4^c) \nonumber\\
    & = \Pp(\msf{H}_{\ell}=H^0 \mid\tilde{\bm{\msf{x}}}_{\ell}=x^\star\cdot\bm{1}) \nonumber\\
    & \leq \varepsilon/4.
\end{align}
By property~1 and \eqref{eq:H1}, and using that $\card{\mc{D}_1} \leq
\card{\mc{D}_2} = \card{\mc{D}_3} = \card{\mc{D}_4} = \dots$,
\begin{align*}
    \Pp_m(\mc{E}_2\mid\mc{E}_3^c\cap\mc{E}_4^c)
    & \leq \sum_{m'\neq m}\sum_{\ell\in\mc{D}_{m'}}
    \Pp_m(\mc{E}_{2,\ell}\mid\mc{E}_3^c\cap\mc{E}_4^c) \\
    & \leq \sum_{m'\neq m}\sum_{\ell\in\mc{D}_{m'}}
    \Pp(\msf{H}_\ell=H^1\mid\tilde{\bm{\msf{x}}}_\ell=\bm{0}) \\
    & \leq M\card{\mc{D}_2} 2^{-B\mu D(W(\cdot|x^\star)||W(\cdot|0)) + o(\log(M))}.
\end{align*}
Now,
\begin{align*}
    \card{\mc{D}_2} \leq 2\nu+1 \leq O(M)
\end{align*}
as $M\to\infty$. Hence, using the definition of $B$,
\begin{align}
    \label{eq:e4}
    \Pp_m(\mc{E}_2\mid\mc{E}_3^c\cap\mc{E}_4^c)
    & \leq 2^{2\log(M) -B\mu D(W(\cdot|x^\star)||W(\cdot|0)) + o(\log(M))} \nonumber\\
    & \leq 2^{-\delta\log(M) + o(\log(M))} \nonumber\\
    & \leq \varepsilon/4
\end{align}
for $M$ large enough.

Substituting \eqref{eq:e1}, \eqref{eq:e2}, \eqref{eq:e3}, and
\eqref{eq:e4} into \eqref{eqn:pe_dmc} shows that for $M$ large enough
the probability of decoding error is upper bounded by $\varepsilon$ for
every message $m$.  By \eqref{eq:power}, the achievable rate per unit
cost of this scheme is
\begin{align*}
    \hat{R}
    & = \frac{\log(M)}{P} \\
    & = \frac{\log(M)}{Bc(x^\star)} \\
    & \geq \frac{\mu}{2+\delta}\frac{D(W(\cdot|x^\star)||W(\cdot|0))}{c(x^\star)}.
\end{align*}
Since $\delta>0$ can be made arbitrarily small, this shows that
\begin{align*}
    \hat{C} \geq \frac{\mu}{2}\frac{D(W(\cdot|x^\star)||W(\cdot|0))}{c(x^\star)}.
\end{align*}
Taking the supremum over all $x^\star\in\mc{X}\setminus \{0\}$
completes the proof of the lower bound in
Theorem~\ref{thm:Chat_DMC}.\hfill\IEEEQED

\begin{remark}
    When using pulse-position modulation over a perfectly synchronized
    channel, the decoder knows exactly where the possible pulses are
    located, and thus needs to check only $M$ possible pulse positions.
    However, in the proposed scheme for channels with synchronization
    errors, {\color{black} the codeword length is set to be of order
    $\Theta(M^2)$ to combat synchronization errors. This translates into
    the number $\card{\mc{D}_m}$ of possible pulse positions for message
    $m$ being $\Theta(M)$. Thus, the sliding window-decoder needs to check
    $M^2$ possible pulse positions due to the lack of synchronization.} It
    is this increase from $M$ to $M^2$ hypothesis tests that results in the
    reduction of rate per unit cost by a factor $2$ compared to the
    synchronized case.
\end{remark}

\section{Proof of Upper Bound in Theorem~\ref{thm:Chat_DMC}}
\label{sec:converse}

In this section, we provide an upper bound on the capacity per unit cost
of the duplication/{\allowbreak}deletion/{\allowbreak}substitution
channel. Since the DDC part of the channel is not memoryless, standard
converse techniques are not applicable in this setting. Instead, we use
a simulation argument, namely that the
duplication/{\allowbreak}deletion/{\allowbreak}substitution channel can be
simulated with an encoder and decoder communicating over a discrete
memoryless channel. This DMC, in turn, can be analyzed and yields the
desired upper bound for capacity per unit cost of the
duplication/{\allowbreak}deletion/{\allowbreak}substitution channel. We
now provide the details of this argument.

Let $f^T$ and $\{\varphi^\ell\}_{\ell=0}^\infty$ be an
encoder-decoder pair achieving rate per unit cost
$\hat{C}(\mu,\sigma^2,W)-\delta$ and average probability of error at
most $\varepsilon$ for the
duplication/{\allowbreak}deletion/{\allowbreak}substitution channel. Note
that, since the output of the channel is of random length $\msf{L}$, the
decoder consists of several sub-decoders
$\{\varphi^\ell\}_{\ell=0}^\infty$, one for each possible realization
$\ell$ of $\msf{L}$.

\begin{figure}[htbp]
    \centering
    \includegraphics{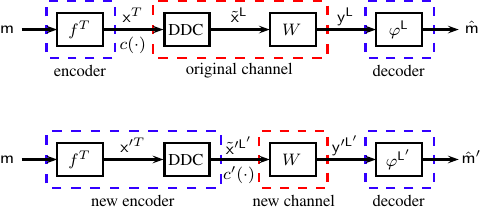}
    \caption{The behavior of the original
    duplication/{\allowbreak}deletion/{\allowbreak}substitution channel
    (top figure) model can be simulated over the discrete memoryless channel
    $W$ by modifying the encoder and the cost function (bottom figure).}
    \label{fig:converse}
\end{figure}

We want to argue that the statistical behavior between the input $m$ to
the encoder $f^T$ and the output $\hat{\msf{m}}$ of the decoder
$\varphi^{\msf{L}}$ can be simulated over the discrete memoryless channel
$W$ alone (see Fig.~\ref{fig:converse}). Consider a new encoder
$\msf{f}'^T$ that consists of the concatenation of $f^T$ with a DDC of
same statistical behavior as the one in the original
duplication/{\allowbreak}deletion/{\allowbreak}substitution channel.
Denote by $\msf{s}'[\ell]$ the state random variables describing this
DDC. Observe that the encoder $\msf{f}'^T$ is a randomized,
variable-length encoder, mapping the message $m$ into a random sequence
$\tilde{\msf{x}}'^{\msf{L}'}$ of random length $\msf{L}'$.

The output $\tilde{\msf{x}}'^{\msf{L}'}$ of the encoder $\msf{f}'^T$ is
transmitted over a DMC with the same transition probability matrix $W$
as in the original
duplication/{\allowbreak}deletion/{\allowbreak}substitution channel. Let
$\msf{y}'^{\msf{L}'}$ be the output of this DMC. The decoder
$\varphi'^{\msf{L}'}$ for the DMC is equal to $\varphi^{\msf{L}'}$.
Observe that this is a variable-length decoder, and denote by
$\hat{\msf{m}}'$ its output.

By construction, for the same message $m$, the distributions of
$\hat{\msf{m}}$ and $\hat{\msf{m}}'$ are identical. In particular, the
average probability of error of both systems is the same. Hence, the
average probability of error of $\msf{f}'^T$ and
$\{\varphi'^{\ell}\}_{\ell=0}^\infty$ over $W$ is at most $\varepsilon$.

Define the new cost function
\begin{equation*}
    c'(\cdot) \defeq \frac{1}{\mu}c(\cdot)
\end{equation*}
for the simulating DMC. With respect to this cost function, and assuming
a uniformly distributed message $\msf{m}$, the expected
cost of using the variable-length encoder $\msf{f}'^T$ over $W$ is
\begin{align*}
    \E c'(\tilde{\msf{x}}'^{\msf{L}'})
    & \stackrel{(a)}{=} \frac{1}{\mu}\E\biggl(
    \sum_{\ell=1}^{\msf{L}'}c(\tilde{\msf{x}}'[\ell]) \biggr)  \\
    & = \frac{1}{\mu}\E\biggl( \sum_{t=1}^{T}\msf{s}'[t]c(\msf{x}'[t]) \biggr)\\
    & \stackrel{(b)}{=} \frac{1}{\mu} \E(\msf{s}'[1])\sum_{t=1}^{T}\E c(\msf{x}'[t]) \\
    & \stackrel{(c)}{=}\frac{1}{\mu} \E(\msf{s}[1]) \sum_{t=1}^{T} \E c(\msf{x}[t]) \\
    & = \E c(\msf{x}^T).
\end{align*}
Here, (a) follows from the definition of $c'(\cdot)$; (b) follows from
the fact that the duplication/{\allowbreak}deletion process $\{\msf{s}'[t]\}$ is
identically distributed and independent of the channel inputs; and (c)
follows since $\msf{s}'[t]$ and $\msf{s}[t]$ have the same
distribution, and since $\msf{x}'[t]$ and $\msf{x}[t]$ have the same
distribution.

Hence, encoder $\msf{f}'^T$ used over the $\DMC(W)$ has the same
expected cost with respect to the cost function $c'(\cdot)$ as the
encoder $f^T$ used over the
duplication/{\allowbreak}deletion/{\allowbreak}substitution channel with
respect to the cost function $c(\cdot)$. Observe that, while the two
encoders have the same \emph{expected} cost, the encoder $f^T$ satisfies
the stronger \emph{per-codeword} cost constraint, whereas the encoder
$\msf{f}'^T$ does not.

The arguments in the last two paragraphs show that there exists a
randomized variable-length encoder $\msf{f}'^T$ and variable-length
decoder $\{\varphi'^{\ell}\}_{\ell=0}^\infty$ achieving a rate per average
unit cost of
\begin{align*}
    \hat{R}'
    & = \frac{\log(M)}{\E c'(\tilde{\msf{x}}'^{\msf{L}'})} \\
    & = \frac{\log(M)}{\E c(\msf{x}^T)} \\
    & \geq \hat{C}(\mu,\sigma^2,W)-\delta
\end{align*}
and average probability of error at most $\varepsilon$. Since this is
just one possible coding scheme, as $M\to\infty$, $\hat{R}'$ must be
upper bounded by $\hat{C}'(W)$, the capacity per unit cost of the
$\DMC(W)$ subject to expected cost constraint, and allowing randomized
variable-length codes. Thus, letting $\delta\to 0$,
\begin{equation}
    \label{eq:upper1}
    \hat{C}(\mu,\sigma^2,W) \leq \hat{C}'(W).
\end{equation}

It remains to analyze $\hat{C}'(W)$. By~\cite[Exercise~6.28]{csiszar11},
we have that for DMCs under expected cost and with \textit{deterministic} variable-length codes
\begin{equation}
    \label{eq:Chat_deterministic}
    \hat{C}'(W) = \max_{\tilde{\msf{x}}'}
    \frac{I(\tilde{\msf{x}}';\msf{y})}{\E c'(\tilde{\msf{x}}')}.
\end{equation}

In Appendix~\ref{apx:random_deterministic}, we show that randomized encoders do not improve capacity per unit cost by arguing that for any randomized encoder, there exists a deterministic encoder that can achieve a not much higher error probability with asymptotically the same cost. This shows that \eqref{eq:Chat_deterministic} is valid for randomized encoders as well.

Note that up to this point we do not require the zero-cost symbol. Now, assuming there is one, we get by \cite[Theorems~2 and 3]{verdu90} that
\begin{align*}
    \max_{\tilde{\msf{x}}'} \frac{I(\tilde{\msf{x}}';\msf{y})}{\E c'(\tilde{\msf{x}}')}
    & = \sup_{\tilde{x}'\in\mc{X}\setminus\{0\}}\frac{D(W(\cdot|\tilde{x}')\Vert W(\cdot|0))}{c'(\tilde{x}')} \\
    & = \mu \sup_{x\in\mc{X}\setminus\{0\}}\frac{D(W(\cdot|x)\Vert W(\cdot|0))}{c(x)},
\end{align*}
so that
\begin{equation}
    \label{eq:upper2}
    \hat{C}'(W) = \mu \sup_{x\in\mc{X}\setminus\{0\}}\frac{D(W(\cdot|x)\Vert W(\cdot|0))}{c(x)}.
\end{equation}
Combining \eqref{eq:upper1} and \eqref{eq:upper2} shows that
\begin{equation*}
    \hat{C}(\mu,\sigma^2,W)
    \leq \mu \sup_{x\in\mc{X}\setminus\{0\}}\frac{D(W(\cdot|x)\Vert W(\cdot|0))}{c(x)},
\end{equation*}
completing the proof.\hfill\IEEEQED

\section{Proof of Theorem~\ref{thm:Chat_Gauss}}
\label{sec:propose_gauss}

The upper bound in Theorem~\ref{thm:Chat_Gauss} follows from the upper
bound for DMCs. Indeed, by Corollary~\ref{cor:Chat_DMC},
\begin{equation*}
    \hat{C}\bigl(\mu,\sigma^2,\mc{N}(0,\eta^2)\bigr)
    \leq \mu\hat{C}\bigl(\mc{N}(0,\eta^2)\bigr)
    = \frac{\mu}{2\eta^2\ln(2)},
\end{equation*}
yielding the desired upper bound.

For the lower bound, we adapt the achievable scheme for the DMC
described in Section~\ref{sec:propose} to the Gaussian case. For
simplicity, we consider the case of noise power $\eta^2=1$ and point out
in the end how to extend the result for arbitrary values of $\eta^2$.

\emph{Encoding:} Fix a target error probability $\varepsilon\in(0,1)$
and a number $\delta\in(0,1)$. Let $x^\star=x^\star(M)>0$ be a nonzero
channel input. Unlike the DMC case, we will choose $x^\star(M)$ as
a function of $M$ here. The codeword for message $m\in\{1,2,\ldots,M\}$ is
again given by
\begin{equation*}
    \bm{x}_m
    \defeq \bigl( \bm{0}_{(m-1)N}, x^\star\cdot \bm{1}_B, \bm{0}_{N-B}, \bm{0}_{(M-m)N} \bigr),
\end{equation*}
with the same
\begin{equation*}
    N \defeq \big\lceil 36M\sigma^2/(\mu^2\varepsilon) \big\rceil
    = \Theta(M)
\end{equation*}
as before. However, here we choose the burst length $B$ as
\begin{equation*}
    B \defeq \big\lfloor \sqrt{MN\sigma^2/\mu^2}\; \big\rfloor
    = \Theta(M),
\end{equation*}
as opposed to $\Theta(\log(M))$ in the DMC case. It can be verified
that $B \leq N$, and hence the codewords are well defined. The block
length of this code is $T=MN$ and the cost of each codeword is
\begin{equation}
    \label{eq:power_b}
    P = B (x^\star)^2.
\end{equation}

\emph{Decoding:} Consider again the subsequences
\begin{align*}
    \bm{\msf{y}}_\ell
    & \defeq \bigl( \msf{y}[\ell], \msf{y}[\ell+1],
    \ldots, \msf{y}[\ell+\floor{B\mu-\beta}-1] \bigr) \\
    \shortintertext{and}\\
    \tilde{\bm{\msf{x}}}_\ell
    & \defeq \bigl( \tilde{\msf{x}}[\ell], \tilde{\msf{x}}[\ell+1],
    \ldots, \tilde{\msf{x}}[\ell+\floor{B\mu-\beta}-1] \bigr)
\end{align*}
of length $\floor{B\mu-\beta}$ for $\ell\in \{ 1, 2, \dots,
\msf{L}-\floor{B\mu-\beta}+1\}$ with
\begin{equation*}
    \beta \defeq \sqrt{4B\sigma^2/\varepsilon}
    = \Theta(\sqrt{M}).
\end{equation*}
Define the open intervals $\tilde{\mc{D}}_m\defeq ((m-1)N\mu+1-\nu,(m-1)N\mu+1+\nu)$ for $m\in\{2,\dots,M\}$. Define the decision regions
\begin{align*}
    \mc{D}_m \defeq
    \begin{cases}
        \{1\}, &\text{for $m=1$} \\
        (\floor{M/\log(M)}\cdot\N) \cap \tilde{\mc{D}}_m, &\text{for $m\in\{2,\dots,M\}$}
    \end{cases}
\end{align*}
with
\begin{equation*}
    \nu \defeq \sqrt{4MN\sigma^2/\varepsilon}
    = \Theta(M).
\end{equation*}
In words, the decision regions consist of $\{1\}$ (for $m=1$) or of all
multiples of $\floor{M/\log(M)}$ between $(m-1)N\mu+1-\nu$ and
$(m-1)N\mu+1+\nu$ (for $m > 1$). This differs from the DMC case, where the
boundaries of the decision regions are the same, but there the regions contain
every \emph{integer} between them. Using the same arguments as in the
DMC case shows that these decision regions are disjoint.

The receiver independently performs the hypothesis test
\begin{equation*}
    \frac{1}{\sqrt{\floor{B\mu-\beta}}}\langle\bm{\msf{y}}_\ell,\bm{1}\rangle
    \mathop{\gtrless}_{H^0}^{H^1} \sqrt{(2+\delta)\ln(M)}
\end{equation*}
for each $\ell\in\mc{D}_m$, $m\in\{1,\dots,M\}$, and where
$\langle\cdot,\cdot\rangle$ denotes the inner product.  Let
$\hat{\msf{H}}_\ell$ be the decision of the hypothesis test for
$\bm{\msf{y}}_\ell$. As in the DMC case, the receiver declares that
message $m$ was sent if $\hat{\msf{H}}_\ell = H^1$ for \emph{some}
$\ell\in\mc{D}_m$ and $\hat{\msf{H}}_\ell = H^0$ for \emph{all}
$\ell\in\mc{D}_{m'}$ with $m'\neq m$. If no such $m$ exists, an error is
declared.

\emph{Error Analysis:} Assume that message $m$ was sent.
We define the same events as in the DMC case. Let
$\mc{E}_{1,\ell}$ be the event that $\hat{\msf{H}}_\ell = H^0$, and
$\mc{E}_{2,\ell}$ be the event that $\hat{\msf{H}}_\ell = H^1$. Define
the missed-detection event
\begin{align*}
    \mc{E}_1 & \defeq \bigcap_{\ell\in\mc{D}_m}\mc{E}_{1,\ell} \\
    \intertext{and the false-alarm event}
    \mc{E}_2 & \defeq \bigcup_{m'\neq m}\bigcup_{\ell\in\mc{D}_{m'}}\mc{E}_{2,\ell}.
\end{align*}
The probability of decoding error for message $m$ is then equal to
$\Pp_m(\mc{E}_1\cup\mc{E}_2)$, where again $\Pp_m$ denotes probability
conditioned on message $m$ being sent.

We again define the two auxiliary events describing the behavior of the
DDC.  Let $\mc{E}_3$ be the event that the total number of symbols in
$\tilde{\bm{\msf{x}}}$ resulting from the first $(m-1)N$ transmitted
symbols is outside $((m-1)N\mu-\nu,(m-1)N\mu+\nu)$, and let $\mc{E}_4$
be the event that the number of symbols in $\tilde{\bm{\msf{x}}}$
resulting from symbols transmitted during time slots $(m-1)N+1$ to
$(m-1)N+B$ is outside $(B\mu-\beta,B\mu+\beta)$. Using the same argument
as for the DMC case, we can upper bound
\begin{align}
    \label{eqn:pe_gauss}
    \Pp_m(\mc{E}_1\cup\mc{E}_2)
    \leq \Pp_m(\mc{E}_3) + \Pp_m(\mc{E}_4) 
    + \Pp_m(\mc{E}_1\mid\mc{E}_3^c \cap \mc{E}_4^c)
    + \Pp_m(\mc{E}_2\mid\mc{E}_3^c \cap \mc{E}_4^c).
\end{align}

Using Chebyshev's inequality as in the analysis of the DMC case, we
obtain
\begin{align}
    \label{eq:e1_b}
    \Pp_m(\mc{E}_3) & \leq \varepsilon/4 \\
    \shortintertext{and} \notag\\
    \label{eq:e2_b}
    \Pp_m(\mc{E}_4) & \leq \varepsilon/4.
\end{align}

We proceed with the analysis of $\Pp_m(\mc{E}_1|\mc{E}_3^c \cap
\mc{E}_4^c)$ and $\Pp_m(\mc{E}_2|\mc{E}_3^c \cap \mc{E}_4^c)$.
Conditioned on message $m$ being sent and on $\mc{E}_3^c\cap\mc{E}_4^c$,
the elements in the decision regions satisfy the following two
properties for $M$ large enough (not depending on $m$):
\begin{enumerate}
    \item For every $\ell\in\mc{D}_{m'}$ with $m'\neq m$, we have
        $\langle\tilde{\bm{\msf{x}}}_\ell,\bm{1}\rangle=0$. Hence,
        \begin{equation*}
            \frac{1}{\sqrt{\floor{B\mu-\beta}}}\langle\bm{\msf{y}}_\ell,\bm{1}\rangle
        \end{equation*}
        is Gaussian with mean zero and variance one.
    \item There exists at least one $\ell\in\mc{D}_m$ such that
        \begin{equation*}
            \langle\tilde{\bm{\msf{x}}}_\ell,\bm{1}\rangle
            \geq x^\star\bigl(\floor{B\mu-\beta}-M/\log(M)\bigr).
        \end{equation*}
        Hence,
        \begin{equation*}
            \frac{1}{\sqrt{\floor{B\mu-\beta}}}\langle\bm{\msf{y}}_\ell,\bm{1}\rangle
        \end{equation*}
        is Gaussian with mean at least
        \begin{equation*}
            x^\star\sqrt{\floor{B\mu-\beta}}\Bigl(1-\frac{M}{\floor{B\mu-\beta}\log(M)}\Bigr)
        \end{equation*}
        and variance one.
\end{enumerate}

The first property follows by the same arguments as for the DMC case,
using that $B \leq \nu/(2\mu)$ and $\beta = o(B(M))$ as $M\to\infty$.
For the second property, note that by the arguments for the DMC case,
there exists at least one $\ell'\in\{1\}$ for $m=1$ or
$\ell'\in\N\cap((m-1)N\mu+1-\nu,(m-1)N\mu+1+\nu)$ for $m > 1$ such that
$\tilde{\bm{\msf{x}}}_{\ell'}=x^\star\cdot\bm{1}$. However, this value of
$\ell'$ may not be a multiple of $\floor{M/\log(M)}$, and hence may not be
an element of $\mc{D}_m$. Let $\ell$ be the closest multiple of
$\floor{M/\log(M)}$ to $\ell'$ that is in $\mc{D}_m$; such a $\ell$
exists for $M$ large enough since $\nu=\Theta(M)$. Since
$\abs{\ell-\ell'}\leq M/\log(M)$, this implies that
\begin{align*}
    \langle\tilde{\bm{\msf{x}}}_\ell,\bm{1}\rangle
    &\geq \langle\tilde{\bm{\msf{x}}}_{\ell'},\bm{1}\rangle-x^\star M/\log(M) \\
    &= x^\star\bigl(\floor{B\mu-\beta}-M/\log(M)\bigr),
\end{align*}
as required.

The two properties allow us to analyze the events $\mc{E}_{1}$ and
$\mc{E}_{2}$. By property~1, and using that $\card{\mc{D}_1} \leq
\card{\mc{D}_2} = \card{\mc{D}_3} = \card{\mc{D}_4} = \dots$,
\begin{align*}
    \Pp_m(\mc{E}_2\mid\mc{E}_3^c\cap\mc{E}_4^c) 
    & \leq \sum_{m'\neq m}\sum_{\ell\in\mc{D}_{m'}}
    \Pp_m(\mc{E}_{2,\ell}\mid\mc{E}_3^c\cap\mc{E}_4^c) \\
    & \leq \sum_{m'\neq m}\sum_{\ell\in\mc{D}_{m'}}
    \Pp_m\biggl(\frac{1}{\sqrt{\floor{B\mu-\beta}}}\langle\bm{\msf{y}}_\ell,\bm{1}\rangle
    \geq \sqrt{(2+\delta)\ln(M)} \biggm\vert \mc{E}_3^c\cap\mc{E}_4^c\biggr) \\
    & \leq M\card{\mc{D}_2} Q\bigl(\sqrt{(2+\delta)\ln(M)}\bigr).
\end{align*}
Using the Chernoff bound $Q(a)\leq\exp(-a^2/2)$ for the $Q$-function, we
have
\begin{equation*}
    Q\bigl(\sqrt{(2+\delta)\ln(M)}\bigr)
    \leq M^{-(1+\delta/2)}.
\end{equation*}
Moreover, since $\nu=\Theta(M)$,
\begin{align*}
    \card{\mc{D}_2}
    \leq \frac{2\nu}{M/\log(M)-1}+1
    \leq O(\log(M))
\end{align*}
as $M\to\infty$. Hence,
\begin{align}
    \label{eq:e4_b}
    \Pp_m(\mc{E}_2\mid\mc{E}_3^c\cap\mc{E}_4^c)
    \leq O(M^{-\delta/2}\log(M))
    \leq \varepsilon/4
\end{align}
for $M$ large enough.

Consider then the value of $\ell\in\mc{D}_m$ guaranteed by property~2.
For this $\ell$,
\begin{align*}
    \Pp_m(\mc{E}_1\mid\mc{E}_3^c\cap\mc{E}_4^c) 
    & \leq \Pp_m(\mc{E}_{1,\ell}\mid\mc{E}_3^c\cap\mc{E}_4^c) \\
    & \leq
    \Pp_m\biggl(\frac{1}{\sqrt{\floor{B\mu-\beta}}}\langle\bm{\msf{y}}_\ell,\bm{1}\rangle
    \leq \sqrt{(2+\delta)\ln(M)} \biggm\vert \mc{E}_3^c\cap\mc{E}_4^c\biggr) \\
    & \leq Q\biggl(x^\star\sqrt{\floor{B\mu-\beta}}\Bigl(1-\frac{M}{\floor{B\mu-\beta}\log(M)}\Bigr) 
    -\sqrt{(2+\delta)\ln(M)}\biggr).
\end{align*}
Recall that $B=\Theta(M)$ and $\beta=\Theta(\sqrt{M})$, so that
\begin{equation*}
    \sqrt{\floor{B\mu-\beta}}\Bigl(1-\frac{M}{\floor{B\mu-\beta}\log(M)}\Bigr)
    = \sqrt{B\mu}(1-o(1))
\end{equation*}
as $M\to\infty$. By choosing
\begin{equation}
    \label{eq:xstar}
    x^\star \defeq (1+\delta)\sqrt{(2+\delta)\ln(M)/(B\mu)},
\end{equation}
we obtain
\begin{equation}
    \label{eq:e3_b}
    \Pp_m(\mc{E}_1\mid\mc{E}_3^c\cap\mc{E}_4^c)
    \leq Q\Bigl( (\delta-o(1))\sqrt{(2+\delta)\ln(M)} \Bigr)
    \leq \varepsilon/4
\end{equation}
for $M$ large enough.

Substituting \eqref{eq:e1_b}, \eqref{eq:e2_b}, \eqref{eq:e4_b}, and
\eqref{eq:e3_b} into \eqref{eqn:pe_gauss} shows that for $M$ large
enough the probability of decoding error is upper bounded by
$\varepsilon$ for every message $m$.  By \eqref{eq:power_b} and
\eqref{eq:xstar}, The power required by this scheme is
\begin{align*}
    P & = B(x^\star)^2 \\
    & = (1+\delta)^2(2+\delta)\ln(M)/\mu.
\end{align*}
Hence, the achievable rate per unit cost for this scheme is
\begin{align*}
    \hat{R}
    & = \frac{\log(M)}{P} \\
    & \geq \frac{\mu}{(1+\delta)^2(2+\delta)\ln(2)}.
\end{align*}
Since $\delta>0$ can be made arbitrarily small, this shows that, for
noise power $\eta^2=1$,
\begin{align*}
    \hat{C} \geq \frac{\mu}{2\ln(2)}.
\end{align*}

Assume then that $\eta^2\neq 1$. By scaling the channel input
at the transmitter by a factor $\eta$ and the channel output at the
receiver by a factor $1/\eta$, we can transform the channel to one with
unit noise power. Since this increases the energy of the transmitted
symbols by a factor $\eta^2$, but does not change the probability of
error, this shows that
\begin{align*}
    \hat{C} \geq \frac{\mu}{2\eta^2\ln(2)},
\end{align*}
concluding the proof. \hfill\IEEEQED

\section{Proof of Theorem~\ref{thm:Chat_compound}}
\label{sec:propose_compound}

This section adapts the coding scheme for the Gaussian
duplication/{\allowbreak}deletion/{\allowbreak}substitution channel with
known value of $\mu$ described in Section~\ref{sec:propose_gauss} to the
compound setting with $\mu$ only known to be in the range
$[\mu_1,\mu_2]$. As before, we will first assume that the noise power is
$\eta^2=1$ and then generalize the result for arbitrary values of
$\eta^2$.

\begin{figure*}[htbp]
    \centering
    \includegraphics{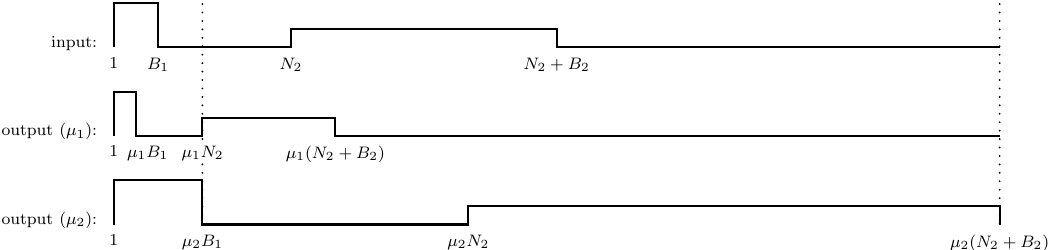}
    \caption{Construction of input waveforms for the Gaussian compound
    duplication/{\allowbreak}deletion/{\allowbreak}substitution channel.
    In the figure, $\mu_1=1/2$ and $\mu_2=2$. For simplicity, $\delta$
    is set to $0$. $N_2$ is chosen such that $\mu_1N_2=\mu_2B_1$,
    ensuring that pulses corresponding to different messages (indicated
    by the dotted lines) are nonoverlapping at the receiver for all
    values of $\mu\in[\mu_1,\mu_2]$ and under expected behavior of the
    DDC.}
    \label{fig:compound}
\end{figure*}

\emph{Encoding:} Fix a target error probability $\varepsilon\in(0,1)$
and a number $\delta\in(0,\mu_1)$. Let
$x^\star=x^\star(M)>0$ be a nonzero channel
input. The codeword for message $m\in\{1,2,\ldots,M\}$ is
\begin{align*}
    \bm{x}_m
    & \defeq \bigl( \bm{0}_{N_m}, x^\star B_m^{-1/2}\cdot \bm{1}_{B_m}, \bm{0}_{T-B_m-N_m} \bigr), \\
    \intertext{where}
    N_m & \defeq
    \begin{cases}
        0, & \text{if $m=1$} \\
        \big\lceil \frac{\mu_2+\delta}{\mu_1-\delta}(N_{m-1}+B_{m-1}) \big\rceil,
        & \text{if $m > 1$}
    \end{cases} \\
    \intertext{and}
    B_m & \defeq
    \begin{cases}
        \floor{\log(M)}, & \text{if $m=1$} \\
        \floor{(\mu_2-\mu_1+2\delta) N_m},
        & \text{if $m > 1$}.
    \end{cases}
\end{align*}
Observe that here, unlike the case with known $\mu$, the value of the
nonzero channel input is $x^\star B_m^{-1/2}$, which depends on the
message $m$. This construction is illustrated in
Fig.~\ref{fig:compound}. The block length of this code is $T = N_M+B_M$,
and the cost of codeword $m$ is
\begin{equation}
    \label{eq:power_c}
    P
    = B_m (x^\star B_m^{-1/2})^2
    = (x^\star)^2.
\end{equation}

\emph{Decoding:} Define the open intervals $\tilde{\mc{D}}_m \defeq ((\mu_1-\delta)N_m+1,(\mu_2+\delta)N_m+1)$ for $m\in\{2,\dots,M\}$ and define the decision regions
\begin{align*}
    \mc{D}_m \defeq
    \begin{cases}
        \{1\}, &\text{for $m=1$} \\
        (\floor{N_m/\log(M)}\cdot\N)\cap \tilde{\mc{D}}_m, &\text{for $m\in\{2,\dots,M\}$}.
    \end{cases}
\end{align*}
Note that, unlike the case with known value of $\mu$, the decision
regions here have increasing width as a function of $m$. However, each decision
region contains approximately the same number
$(\mu_2-\mu_1+2\delta)\log(M)$ of points. It is easy to verify that the
decision regions are disjoint.

For $\ell\in\mc{D}_m$, define the subsequences
\begin{align*}
    \bm{\msf{y}}_\ell
    & \defeq \bigl( \msf{y}[\ell], \msf{y}[\ell+1], \ldots,
    \msf{y}[\ell+\floor{(\mu_1-\delta)B_m}-1] \bigr) \\
    \shortintertext{and}\\
    \tilde{\bm{\msf{x}}}_\ell
    & \defeq \bigl( \tilde{\msf{x}}[\ell], \tilde{\msf{x}}[\ell+1],
    \ldots, \tilde{\msf{x}}[\ell+\floor{(\mu_1-\delta)B_m}-1] \bigr)
\end{align*}
of length $\floor{(\mu_1-\delta)B_m}$. We point out that here, unlike
the case with known value of $\mu$, the subsequences in different
regions $\mc{D}_{m}$ and $\mc{D}_{m'}$ have different lengths.

The receiver independently performs the hypothesis test
\begin{equation*}
    \frac{1}{\sqrt{\floor{(\mu_1-\delta)B_m}}}\langle\bm{\msf{y}}_\ell,\bm{1}\rangle
    \mathop{\gtrless}_{H^0}^{H^1} \sqrt{(2+\delta)\ln(M)}
\end{equation*}
for each $\ell\in\mc{D}_m$, $m\in\{1,\dots,M\}$, and where
$\langle\cdot,\cdot\rangle$ denotes the inner product.  Let
$\hat{\msf{H}}_\ell$ be the decision of the hypothesis test for
$\bm{\msf{y}}_\ell$. As in the case of known $\mu$, the receiver
declares that message $m$ was sent if $\hat{\msf{H}}_\ell = H^1$ for
\emph{some} $\ell\in\mc{D}_m$ and $\hat{\msf{H}}_\ell = H^0$ for
\emph{all} $\ell\in\mc{D}_{m'}$ with $m'\neq m$. If no such $m$ exists,
an error is declared.

\emph{Error Analysis:} Assume that message $m$ was sent.  We define the
same error events as in the case of known $\mu$. Let $\mc{E}_{1,\ell}$
be the event that $\hat{\msf{H}}_\ell = H^0$, and $\mc{E}_{2,\ell}$ be
the event that $\hat{\msf{H}}_\ell = H^1$. Define the missed-detection
event
\begin{align*}
    \mc{E}_1 & \defeq \bigcap_{\ell\in\mc{D}_m}\mc{E}_{1,\ell} \\
    \shortintertext{and the false-alarm event}\\
    \mc{E}_2 & \defeq \bigcup_{m'\neq m}\bigcup_{\ell\in\mc{D}_{m'}}\mc{E}_{2,\ell}.
\end{align*}
The probability of decoding error for message $m$ is then equal to
$\Pp_m(\mc{E}_1\cup\mc{E}_2)$, where as before $\Pp_m$ denotes
probability conditioned on message $m$ being sent.

We again define two auxiliary events describing the behavior of the
DDC. Let $\mc{E}_3$ be the event that the total number of symbols in
$\tilde{\bm{\msf{x}}}$ resulting from the first $N_m$ transmitted
symbols is outside $((\mu_1-\delta)N_m,(\mu_2+\delta)N_m)$, and
let $\mc{E}_4$ be the event that the number of symbols in
$\tilde{\bm{\msf{x}}}$ resulting from symbols transmitted during time
slots $N_m+1$ to $N_m+B_m$ is outside
$((\mu_1-\delta)B_m,(\mu_2+\delta)B_m)$. We can again upper bound the
probability of error as
\begin{align}
    \label{eqn:pe_compound}
    \Pp_m(\mc{E}_1\cup\mc{E}_2)
    \leq \Pp_m(\mc{E}_3) + \Pp_m(\mc{E}_4) 
    + \Pp_m(\mc{E}_1\mid\mc{E}_3^c \cap \mc{E}_4^c)
    + \Pp_m(\mc{E}_2\mid\mc{E}_3^c \cap \mc{E}_4^c).
\end{align}

We start with the analysis of $\Pp_m(\mc{E}_3)$ and $\Pp_m(\mc{E}_4)$.  Using
Chebyshev's inequality together with the upper bound $\sigma^2$ on the
variance of the states $\msf{s}[t]$, we obtain similarly to the case
with known value of $\mu$
\begin{align}
    \label{eq:e1_c}
    \Pp_m(\mc{E}_3) & \leq \varepsilon/4
\end{align}
and
\begin{align}
    \label{eq:e2_c}
    \Pp_m(\mc{E}_4) & \leq \varepsilon/4
\end{align}
for $M$ large enough and for any value of $\mu\in[\mu_1,\mu_2]$.

We proceed with the analysis of $\Pp_m(\mc{E}_1|\mc{E}_3^c \cap
\mc{E}_4^c)$ and $\Pp_m(\mc{E}_2|\mc{E}_3^c \cap \mc{E}_4^c)$.
Conditioned on message $m$ being sent and on $\mc{E}_3^c\cap\mc{E}_4^c$,
the elements in the decision regions satisfy the following two
properties for $M$ large enough (not depending on $m$):
\begin{enumerate}
    \item For every $\ell\in\mc{D}_{m'}$ with $m'\neq m$, we have
        $\langle\tilde{\bm{\msf{x}}}_\ell,\bm{1}\rangle=0$. Hence,
        \begin{equation*}
            \frac{1}{\sqrt{\floor{(\mu_1-\delta)B_m}}}\langle\bm{\msf{y}}_\ell,\bm{1}\rangle
        \end{equation*}
        is Gaussian with mean zero and variance one.
    \item There exists at least one $\ell\in\mc{D}_m$ such that
        \begin{equation*}
            \langle\tilde{\bm{\msf{x}}}_\ell,\bm{1}\rangle
            \geq x^\star B_m^{-1/2}\bigl( \floor{(\mu_1-\delta)B_m}-N_m/\log(M)\bigr).
        \end{equation*}
        Hence,
        \begin{equation*}
            \frac{1}{\sqrt{\floor{(\mu_1-\delta)B_m}}}\langle\bm{\msf{y}}_\ell,\bm{1}\rangle
        \end{equation*}
        is Gaussian with mean at least
        \begin{equation*}
            x^\star\sqrt{(\mu_1-\delta-1/B_m)}
            \Bigl(1-\frac{N_m}{\floor{(\mu_1-\delta)B_m}\log(M)}\Bigr)
        \end{equation*}
        and variance one.
\end{enumerate}

Property~2 can be proved using arguments analogous to the case with
known value of $\mu$. For property~1, we need to argue that the burst of
symbols $x^\star$ cannot be shifted into the incorrect decoding region.

Assume first $m'< m$. The right-most element of $\mc{D}_{m'}$ is at
position less than or equal to $(\mu_2+\delta)N_{m-1}+1$,
and thus the right-most element of $\tilde{\bm{\msf{x}}}_\ell$ with
$\ell\in\mc{D}_{m'}$ is at position at most
\begin{equation*}
    (\mu_2+\delta)N_{m-1}+(\mu_1-\delta)B_{m-1}
    \leq (\mu_2+\delta)(N_{m-1}+B_{m-1}).
\end{equation*}
Now, conditioned on $\mc{E}_3^c$, there are at least
$(\mu_1-\delta)N_m$ symbols $0$ in $\tilde{\bm{\msf{x}}}$ before the
first symbol $x^\star$. For there to be no overlap, it is sufficient to
argue that
\begin{equation*}
    (\mu_2+\delta)(N_{m-1}+B_{m-1})
    \leq (\mu_1-\delta)N_m
\end{equation*}
or, equivalently, that
\begin{equation*}
    N_m \geq \frac{\mu_2+\delta}{\mu_1-\delta}(N_{m-1}+B_{m-1}).
\end{equation*}
This holds by the definition of $N_m$.

Assume then that $m' > m$. The left-most element of any
$\tilde{\bm{\msf{x}}}_\ell$ with $\ell\in\mc{D}_{m'}$ is at position at
least $(\mu_1-\delta)N_{m+1}+1$. Conditioned on $\mc{E}_3^c$, there are
at most $(\mu_2+\delta)N_m$ symbols $0$ before the first symbol
$x^\star$ in $\tilde{\bm{\msf{x}}}$. Conditioned on $\mc{E}_4^c$, the
burst of symbol $x^\star$ in $\tilde{\bm{\msf{x}}}$ is of length at most
$(\mu_2+\delta)B_m$.  For there to be no overlap, it is sufficient that
\begin{equation*}
    (\mu_2+\delta)N_m+(\mu_2+\delta)B_m \leq (\mu_1-\delta)N_{m+1},
\end{equation*}
or, equivalently, that
\begin{equation*}
    N_{m+1} \geq \frac{\mu_2+\delta}{\mu_1-\delta}(N_m+B_m).
\end{equation*}
This holds again by the definition of $N_{m+1}$.

The two properties allow us to analyze the events $\mc{E}_{1}$ and
$\mc{E}_{2}$. By property~1,
\begin{align*}
    \Pp_m(\mc{E}_2\mid\mc{E}_3^c\cap\mc{E}_4^c) 
    & \leq \sum_{m'\neq m}\sum_{\ell\in\mc{D}_{m'}}
    \Pp_m(\mc{E}_{2,\ell}\mid\mc{E}_3^c\cap\mc{E}_4^c) \\
    & \leq \sum_{m'\neq m}\sum_{\ell\in\mc{D}_{m'}} \Pp_m\biggl(\frac{1}{\sqrt{\floor{(\mu_1-\delta)B_m}}}\langle\bm{\msf{y}}_\ell,\bm{1}\rangle 
    \geq \sqrt{(2+\delta)\ln(M)} \biggm\vert \mc{E}_3^c\cap\mc{E}_4^c\biggr) \\
    & \leq Q\bigl( \sqrt{(2+\delta)\ln(M)} \bigr)\sum_{m'=1}^M\card{\mc{D}_{m'}}.
\end{align*}
Using the Chernoff bound $Q(a)\leq\exp(-a^2/2)$ for the $Q$-function,
\begin{equation*}
    Q\bigl(\sqrt{(2+\delta)\ln(M)}\bigr)
    \leq M^{-(1+\delta/2)}.
\end{equation*}
Moreover,
\begin{align*}
    \card{\mc{D}_{m'}}
    \leq \frac{(\mu_2-\mu_1+2\delta)N_m}{N_m/\log(M)-1}+1
    \leq O(\log(M))
\end{align*}
so that
\begin{equation*}
    \sum_{m'=1}^M\card{\mc{D}_{m'}}
    \leq O(M\log(M))
\end{equation*}
as $M\to\infty$. Hence,
\begin{align}
    \label{eq:e4_c}
    \Pp_m(\mc{E}_2\mid\mc{E}_3^c\cap\mc{E}_4^c)
    \leq O(M^{-\delta/2}\log(M))
    \leq \varepsilon/4
\end{align}
for $M$ large enough.

Consider then the value of $\ell\in\mc{D}_m$ guaranteed by property~2. For this $\ell$,
\begin{align*}
    \Pp_m(\mc{E}_1\mid\mc{E}_3^c\cap\mc{E}_4^c) 
    & \leq \Pp_m(\mc{E}_{1,\ell}\mid\mc{E}_3^c\cap\mc{E}_4^c) \\
    & \leq
    \Pp_m\biggl(\frac{\langle\bm{\msf{y}}_\ell,\bm{1}\rangle}{\sqrt{\floor{(\mu_1-\delta)B_m}}} 
    \leq \sqrt{(2+\delta)\ln(M)} \biggm\vert \mc{E}_3^c\cap\mc{E}_4^c\biggr) \\
    &\leq Q\biggl(x^\star\sqrt{(\mu_1-\delta-1/B_m)}\Bigl(1-\frac{N_m}{\floor{(\mu_1-\delta)B_m}\log(M)}\Bigr) 
    -\sqrt{(2+\delta)\ln(M)}\biggr).
\end{align*}
Note that
\begin{equation*}
    1-\frac{N_m}{\floor{(\mu_1-\delta)B_m}\log(M)}
    = 1
\end{equation*}
for $m=1$, and
\begin{align*}
    1-\frac{N_m}{\floor{(\mu_1-\delta)B_m}\log(M)} 
    \geq 1-\frac{1}{\bigl((\mu_1-\delta)(\mu_2-\mu_1+2\delta-1/N_m)-1/N_m\bigr)\log(M)} 
    \geq 1-o(1)
\end{align*}
as $M\to\infty$ for $m > 1$. Furthermore,
\begin{equation*}
    \sqrt{(\mu_1-\delta-1/B_m)}
    \geq \sqrt{(\mu_1-\delta)}(1-o(1))
\end{equation*}
as $M\to\infty$.

By choosing
\begin{equation}
    \label{eq:xstar_c}
    x^\star \defeq (1+\delta)\sqrt{(2+\delta)\ln(M)/(\mu_1-\delta)},
\end{equation}
we obtain
\begin{align}
    \label{eq:e3_c}
    \Pp_m(\mc{E}_1\mid\mc{E}_3^c\cap\mc{E}_4^c)
    \leq Q\Bigl( (\delta-o(1))\sqrt{(2+\delta)\ln(M)} \Bigr) 
    \leq \varepsilon/4
\end{align}
for $M$ large enough.

Substituting \eqref{eq:e1_c}, \eqref{eq:e2_c}, \eqref{eq:e4_c}, and
\eqref{eq:e3_c} into \eqref{eqn:pe_compound} shows that for $M$ large
enough the probability of decoding error is upper bounded by
$\varepsilon$ for every message $m$. By \eqref{eq:power_c} and
\eqref{eq:xstar_c}, the power required by this scheme is
\begin{align*}
    P = (x^\star)^2 
    = (1+\delta)^2(2+\delta)\ln(M)/(\mu_1-\delta).
\end{align*}
Hence, the achievable rate per unit cost for this scheme is
\begin{align*}
    \hat{R}
    & = \frac{\log(M)}{P} \\
    & \geq \frac{\mu_1-\delta}{(1+\delta)^2(2+\delta)\ln(2)}.
\end{align*}
Since $\delta>0$ can be made arbitrarily small, this shows that, for
noise power $\eta^2=1$,
\begin{align*}
    \hat{C} \geq \frac{\mu_1}{2\ln(2)}.
\end{align*}
By scaling the input and output as in the proof of
Theorem~\ref{thm:Chat_Gauss} in Section~\ref{sec:propose_gauss},
this implies that
\begin{align*}
    \hat{C} \geq \frac{\mu_1}{2\eta^2\ln(2)}
\end{align*}
for any value of noise power $\eta^2$, concluding the proof.  \hfill\IEEEQED

\appendices
\section{Randomization Does Not Increase $\hat{C}'(W)$}\label{apx:random_deterministic}

In this appendix, we show that randomized encoders do not increase
capacity per unit cost. We particularly argue that for any randomized
encoder with probability of error $\bar{\varepsilon}$ and cost
$\bar{c}'$, there exists a deterministic encoder that has a probability
of error upper-bounded by $\sqrt{\bar{\varepsilon}}$, with a cost no
bigger than $\bar{c}'/(1-\sqrt{\bar{\varepsilon}})$. Since
$\sqrt{\bar{\varepsilon}}\rightarrow 0$ and
$\bar{c}'/(1-\sqrt{\bar{\varepsilon}})\rightarrow \bar{c}'$
as $\bar{\varepsilon}\rightarrow 0$, this shows that randomized encoders
and deterministic encoders achieve the same capacity per unit cost.

Consider a randomized encoder $\msf{f}^{'}$. We first point out that
every realization $f^{'}$ of $\msf{f}^{'}$ (corresponding to a
deterministic encoder) has the same message rate $\log(M)$. Let
$\varepsilon(f^{'})$ and $c'(f^{'})$ be the average (over the
codebook) probability of error and average cost for a particular
realization $\msf{f}^{'}=f^{'}$, respectively. The overall probability
of error and cost of this randomized encoder are then
\begin{align*}
    \bar{\varepsilon} &= \mbb{E}(\varepsilon(\msf{f}^{'})) \\
    \shortintertext{and}\\
    \bar{c}' &= \mbb{E}(c'(\msf{f}^{'})),
\end{align*}
respectively. Now, by Carath\'{e}odory's theorem (see for example
\cite[Theorem 17.1]{rockafellar97}), any tuple $(\bar{\varepsilon} ,
\bar{c}')$ achieved by a randomized encoder can be achieved by a convex
combination of at most three deterministic encoders without loss of
generality. Moreover, every point $(\bar{\varepsilon} , \bar{c}')$ in
the interior of the convex hull of these three points corresponding to
the deterministic encoders is dominated by a point on a face of this
convex hull. Therefore, in the following, we can assume without loss of
generality that $\msf{f}^{'}$ is a convex combination of two
deterministic encoders $f^{'}_1$ and $f^{'}_2$ with weights
$0\leq\lambda\leq 1$ and $1-\lambda$, respectively. We thereby have
\begin{align}
    \bar{\varepsilon} &= \lambda\varepsilon(f^{'}_1) + (1-\lambda)\varepsilon(f^{'}_2) \label{eq:e_bar} \\
    \shortintertext{and} \nonumber \\
    \bar{c}' &= \lambda c'(f^{'}_1) + (1-\lambda) c'(f^{'}_2). \label{eq:c_bar}
\end{align}

Now, if either $f^{'}_i$ for $i\in\{1,2\}$ satisfies
$\varepsilon(f^{'}_i)\leq \bar{\varepsilon}$ and
$c'(f^{'}_i)\leq\bar{c}'$, then this $f^{'}_i$ is what we are looking
for and we are done. So we can assume in the following that
\begin{enumerate}
    \item $\varepsilon(f^{'}_1)\leq \bar{\varepsilon}$ and $c'(f^{'}_1)\geq\bar{c}'$, and
    \item $\varepsilon(f^{'}_2)\geq \bar{\varepsilon}$ and $c'(f^{'}_2)\leq\bar{c}'$.
\end{enumerate}
Assume that $\varepsilon(f^{'}_2)\geq \sqrt{\bar{\varepsilon}}$;
otherwise, $\varepsilon(f^{'}_2)<\sqrt{\bar{\varepsilon}}$ and the
deterministic encoder $f^{'}_2$ serves the purpose. Then, from
\eqref{eq:e_bar}, we have
\begin{align*}
    \bar{\varepsilon} &\geq \lambda \varepsilon(f^{'}_1) + (1-\lambda) \sqrt{\bar{\varepsilon}} \\
    &= \lambda\left(\varepsilon(f^{'}_1) - \sqrt{\bar{\varepsilon}}\right) + \sqrt{\bar{\varepsilon}},
\end{align*}
and hence
\begin{equation*}
    \lambda \geq \frac{\sqrt{\bar{\varepsilon}}-\bar{\varepsilon}}{\sqrt{\bar{\varepsilon}}-\varepsilon(f^{'}_1)}.
\end{equation*}
Plugging this bound into \eqref{eq:c_bar} results in
\begin{align*}
    c'(f^{'}_1) 
    &= \frac{1}{\lambda} \left( \bar{c}' - (1-\lambda) c'(f^{'}_2) \right) \\
    & \leq \bar{c}'/\lambda \\
    &\leq \bar{c}' \frac{\sqrt{\bar{\varepsilon}}-\varepsilon(f^{'}_1)}{\sqrt{\bar{\varepsilon}}-\bar{\varepsilon}} \\
    &\leq \frac{\bar{c}'}{1-\sqrt{\bar{\varepsilon}}}.
\end{align*}
Thus, the deterministic encoder $f^{'}_1$ satisfies
$\varepsilon(f^{'}_1)\leq \bar{\varepsilon}$ and $c'(f^{'}_1)\leq
\bar{c}'/(1-\sqrt{\bar{\varepsilon}})$, completing the proof.
\hfill\IEEEQED


\begin{thebibliography}{10}

\bibitem{proakis03}
J.~G. Proakis, {\em Digital Communications}.
\newblock McGraw-Hill, fourth~ed., 2001.

\bibitem{dobrushin67}
R.~L. Dobrushin, ``Shannon's theorems for channels with synchronization
  errors,'' {\em Problems Inform. Transm.}, vol.~3, no.~4, pp.~11--26, 1967.

\bibitem{golay49}
M.~J.~E. Golay, ``Note on the theoretical efficiency of information reception
  using {PPM},'' {\em Proc. IRE}, vol.~37, p.~1031, Sept. 1949.

\bibitem{gallager88}
R.~Gallager, ``Energy limited channels: Coding, multiaccess, and spread
  spectrum,'' in {\em Proc. Conf. Inform. Sci. Syst.}, Mar. 1988.

\bibitem{verdu90}
S.~Verd{\'u}, ``On channel capacity per unit cost,'' {\em IEEE Trans. Inf.
  Theory}, vol.~36, pp.~1019--1030, Sept. 1990.

\bibitem{diggavi06}
S.~Diggavi and M.~Grossglauser, ``On information transmission over a finite
  buffer channel,'' {\em IEEE Trans. Inf. Theory}, vol.~52, pp.~1126--1237,
  Mar. 2006.

\bibitem{mitzenmacher06}
M.~Mitzenmacher and E.~Drinea, ``A simple lower bound for the capacity of the
  deletion channel,'' {\em IEEE Trans. Inf. Theory}, vol.~52, pp.~4657--4660,
  Oct. 2006.

\bibitem{diggavi07}
S.~Diggavi, M.~Mitzenmacher, and H.~Pfister, ``Capacity upper bounds for the
  deletion channel,'' in {\em Proc. IEEE ISIT}, pp.~1716--1720, June 2007.

\bibitem{drinea07}
E.~Drinea and M.~Mitzenmacher, ``Improved lower bounds for the capacity of
  i.i.d. deletion and duplication channels,'' {\em IEEE Trans. Inf. Theory},
  vol.~53, pp.~2693--2714, Aug. 2007.

\bibitem{kirsch10}
A.~Kirsch and E.~Drinea, ``Directly lower bounding the information capacity for
  channels with i.i.d. deletions and duplications,'' {\em IEEE Trans. Inf.
  Theory}, vol.~56, pp.~86--102, Jan. 2010.

\bibitem{fertonani10}
D.~Fertonani and T.~M. Duman, ``Novel bounds on the capacity of the binary
  deletion channel,'' {\em IEEE Trans. Inf. Theory}, vol.~56, pp.~2753--2765,
  June 2010.

\bibitem{kanoria10}
Y.~Kanoria and A.~Montanari, ``On the deletion channel with small deletion
  probability,'' in {\em Proc. IEEE ISIT}, pp.~1002--1006, June 2010.

\bibitem{kalai10}
A.~Kalai, M.~Mitzenmacher, and M.~Sudan, ``Tight asymptotic bounds for the
  deletion channel with small deletion probabilities,'' in {\em Proc. IEEE
  ISIT}, pp.~997--1001, June 2010.

\bibitem{massey72}
J.~Massey, ``Optimum frame synchronization,'' {\em IEEE Trans. Commun.},
  vol.~20, pp.~115--119, Apr. 1972.

\bibitem{chandar13}
V.~Chandar, A.~Tchamkerten, and D.~Tse, ``Asynchronous capacity per unit
  cost,'' {\em IEEE Trans. Inf. Theory}, vol.~59, pp.~1213--1226, Mar. 2013.

\bibitem{bladsjo13}
D.~{Bladsj\"{o}}, M.~Hogan, and S.~Ruffini, ``Synchronization aspects in {LTE}
  small cells,'' {\em IEEE Commun. Mag.}, vol.~51, pp.~70--77, Sept. 2013.

\bibitem{gankumsri03}
S.~Ganeriwal, R.~Kumar, and M.~B. Srivastava, ``Timing-sync protocol for sensor
networks,'' in {\em Proc. ACM SenSys}, pp.~138--149, Nov. 2003.

\bibitem{cover91}
T.~M. Cover and J.~A. Thomas, {\em Elements of Information Theory}.
\newblock Wiley, 1991.

\bibitem{csiszar11}
I.~Csisz{\'a}r and J.~K{\"o}rner, {\em Information Theory: Coding Theorems for
  Discrete Memoryless Systems}.
\newblock Cambridge University Press, second~ed., 2011.

\bibitem{rockafellar97}
R.~T. Rockafellar, {\em Convex Analysis}.
\newblock Princeton University Press, 1997.

\end{thebibliography}
\end{document}